\documentclass[a4paper,11pt]{article}
\usepackage{graphicx}

\usepackage{jcappub} 
\usepackage{mathrsfs}
\usepackage[T1]{fontenc} 
\usepackage{bbold}
\usepackage{url}
\usepackage{hyperref}
\usepackage[english]{babel}
\usepackage[utf8]{inputenc}
\usepackage{amsmath}
\usepackage{color}
\usepackage{amsfonts}
\usepackage{amssymb}
\usepackage{mathtools}
\usepackage{placeins}
\usepackage{float}
\usepackage{tabularx}
\usepackage{adjustbox}
\usepackage{caption}
\usepackage{subcaption}
\usepackage{graphicx}
\usepackage{color}
\usepackage{xcolor, soul}
\sethlcolor{yellow}
\usepackage{amssymb}
\usepackage{amsmath}
\usepackage{array}
\usepackage{mathtools}
\usepackage{placeins}
\usepackage{float}
\usepackage{tabularx}
\usepackage{adjustbox}
\usepackage{xurl}
\usepackage{hyperref}
\usepackage{url}
\usepackage{natbib}
\usepackage[normalem]{ulem}
\usepackage{afterpage}
\usepackage{float}
\usepackage{ulem}
\usepackage{tensor}
\usepackage[T1]{fontenc}
\usepackage[utf8]{inputenc}
\usepackage{graphicx}

\usepackage{mathrsfs}
\usepackage[T1]{fontenc} 
\usepackage{bbold}
\usepackage{url}
\usepackage[english]{babel}
\usepackage[utf8]{inputenc}
\usepackage{amsmath}

\usepackage{amsfonts}
\usepackage{amssymb}
\usepackage{mathtools}
\usepackage{placeins}
\usepackage{float}
\usepackage{tabularx}
\usepackage{adjustbox}
\usepackage{caption}
\usepackage{subcaption}
\DeclareUnicodeCharacter{2212}{-}

\begin{document}
\title{\boldmath Spontaneous baryogenesis and generation of gravitational waves in a new model of quintessential \texorpdfstring{$\alpha$}{a}-attractor}
\author{Arunoday Sarkar}
\author{and Buddhadeb Ghosh}
\affiliation{Centre of Advanced Studies, Department of Physics, The University of Burdwan,\\Burdwan 713 104, India}
\emailAdd{adsarkar@scholar.buruniv.ac.in}
\emailAdd{bghosh@phys.buruniv.ac.in}




\date{\today}

\abstract{We study the role of $\alpha$-parameter of the newly proposed model of quintessential $\alpha$-attractor inflation (arXiv: 2305.00230 [gr-qc]) to the case of quintessential spontaneous baryogenesis and generation of relic gravitational waves in presence of a rolling scalar field during kination. An \textit{effective 4-Fermi construct} technique has been employed to compute the freeze-out temperature and the baryon-to-entropy ratio, of which the obtained results conform to the experimental requirements for $0.28\leq\alpha\leq 0.30$. This range of $\alpha$ is found to originate from the functional behaviour of the end-value expression of the potential concerned. We also find a blue-tilted gravitational wave spectrum during a transition from inflation to kination. The amplitudes of the gravitational waves during radiation domination satisfy the constraint for nucleosynthesis and the characteristic strain of the ongoing gravitational wave detectors. Thus, the most important observation emerged from the present study is that, increasingly small fractional values of $\alpha$ are favourable for unification of inflation, baryogenesis, quintessence and gravitational waves within a single model. This could have an interesting connection with the fundamental origin of $\alpha$-attractor.} 
\maketitle
\flushbottom


\section{Introduction}
\label{sec:intro}
The \textit{matter-antimatter asymmetry} is an intriguing phenomenon in the frontiers of particle physics and cosmology. There is no known exact mechanism to generate the tiny baryon-to-entropy ratio $\eta\approx (8.6\pm 0.1)\times 10^{-11}$ \cite{Cline:2006ts,AharonyShapira:2021ize,Rubakov:2019lyf,DeSimone:2016ofp}, which is a crucial requirement for big bang nucleosynthesis (BBN) and is almost fixed from the early universe to the present day. One can imagine that, due to some exotic process in the early universe, baryon number conservation was violated, which led to the excess number densities of particles over those of anti-particles. This process is popularly known as \textit{Baryogenesis}\footnote{For the related phenomenon, \textit{Leptogenesis} see Ref. \cite{Alexander:2004us}, for example.}. In 1967, Andrei Sakharov proposed three criteria \cite{Sakharov:1967dj}, based on symmetry principles, in order to achieve the baryon asymmetry: i) Baryon number ($B$) violation, ii) Charge conjugation ($C$) and charge conjugation-parity ($CP$) violations and iii) thermal non-equilibrium in presence of charge conjugation-parity-time reversal ($CPT$) invariance. The first condition is trivial in order to achieve a baryon asymmetric universe with $\Delta B\neq 0$. Since any stipulation of $C$ and $CP$ invariance would be compatible with the $B$ invariance, $C$ and $CP$ violations are essentially needed to ensure $B$ violation \cite{Quiros:1999jp}. This is Sakharov's second criterion. Now, to maintain the excess of baryons the rate of $B$ violation should be very small compared to the expansion rate of the universe, otherwise the $CPT$ invariance and hence the $T$ violation would bring back the universe in a state of $\Delta B=0$, through thermal equilibrium. It is only possible by the `\emph{departure from thermal equilibrium}', which is Sakharov's third condition. The last condition is important because in a phase of matter-antimatter symmetry, no preferred direction of time can be made, but thermal out-of equilibrium sets a unique direction of flow of time from the early universe to the present day. \par The Standard Model (SM) of particle physics can accommodate large $B$ violation \cite{Arnold:1987mh} at high temperature by sphaleron transition \cite{Klinkhamer:1984di} between the degenerate $SU(2)$ vacua and very small CP violation \cite{Quiros:1999jp}. However, it can not meet Sakharov's third criterion, as this requires a \emph{Strong First Order Electroweak Phase Transition} (SFOEWPT) \cite{Ghosh:2015rsa,Trodden:2003yn,Aziz:2009hk,Aziz:2010ja}, which is not possible to achieve in SM \cite{Ilgenfritz:1995sh,Gurtler:1996wx,Kajantie:1996qd} for experimentally favoured Higgs mass \cite{LEPWorkingGroupforHiggsbosonsearches:2003ing,DELPHI:2003uey}.\par This was already evident in the large value ($\sim 10^{12}$) \cite{Arnold:1987mh} of the ratio, $B$ violation rate/universe expansion rate, calculated in the SM. However, a small value of this ratio consistent with the SFOEWPT can be obtained in a number of \emph{Beyond Standard Models} (BSMs) (for a review, see Ref. \cite{Ghosh:2015rsa}), at the TeV scale.\par Another angle of comparison between the SM and the BSMs is the fact that in the former the Higgs mass is unstable against the bosonic and the fermionic loop corrections, whereas in the latter it is not, due to the cancellation effects. These theoretical findings make the situation somewhat confusing because the experimental Higgs (mass $\sim 126$ GeV) in the LHC experiments \cite{ATLAS:2012yve,CMS:2012qbp} is SM-like. Only future collider experiments might resolve some of these issues by either proving or disproving the BSMs.\par An interesting alternative approach is to relax Sakharov's third criterion by invoking an effective formulation beyond conventional particle physics models in order to generate the required baryon asymmetry in thermal equilibrium \textit{\`{a} la} \textit{spontaneous baryogenesis} (SB) \cite{Cohen:1987vi,Cohen:1988kt,Cohen:1991iu,Abel:1992za,Reina:1993ws,Comelli:1994di,McDonald:1995np,Comelli:1995uz,Dolgov:1997qr,Takahashi:2003db,Brandenberger:2003kc,Feng:2006dht,Takahashi:2015ula,DeSimone:2016ofp,DeSimone:2016juo,Arbuzova:2016qfh,Dasgupta:2018eha,Grzadkowski:2018nbc,Barenboim:2019vmu,Brandenberger:2019jfh,Domcke:2020kcp,Luongo:2021gho,Foster:2022ajl}. This approach is highly non-trivial because, generically, the average value of baryon number density in thermal equilibrium is zero. However, a non-zero value can be obtained even in thermal equilibrium, by the spontaneous breaking of the CPT invariance \cite{Cohen:1987vi}. This is possible if there exists a rolling scalar field $\phi$ and the associated kinetic energy is responsible for the spontaneous CPT non-invariance. Particularly, it is seen to manifest through the derivative coupling of the $\phi$ field with a non-conserving baryon current $J^{\mu}_b$ as $\partial_\mu\phi J^\mu_b$ \cite{DeSimone:2016ofp,Takahashi:2003db}. Here, the $\phi$ field acts as a pseudo-Nambu-Goldstone boson \cite{Frieman:1995pm,Dolgov:1996qq} in an approximate $U(1)$ symmetry \cite{Cohen:1987vi,Dolgov:1994zq,Dolgov:1996qq}. (See for related studies in particle physics, \cite{Cohen:1991iu,Carroll:2005dj} and in inflationary cosmology, \cite{Dolgov:1994zq,Dolgov:1996qq,Li:2001st}.)\par In the interface of particle physics and cosmology, this scenario is popularly known as \textit{quintessential baryogenesis} \cite{DeFelice:2002ir,Bettoni:2018utf,Hashiba:2019mzm,Ahmad:2019jbm,Jaman:2020srx,Basak:2021cgk} in the paradigm of inflationary quintessence \cite{Peebles:1998qn}, since the idea of rolling scalar field is inbuilt in this kind of inflationary framework. Quite interestingly, here, the scalar field playing the roles of inflaton over the slow-roll plateau and quintessence in the potential runaway, can also trigger the process of baryogenesis during kination. Unlike the ordinary inflationary models, the kination regime makes the quintessential models unique in suitably embedding the SB scenario to explain the observed baryon asymmetry and the generation of relic gravitational waves. The quintessential inflaton field when traverses through the kination regime \cite{Joyce:1996cp} after inflation, particle production starts to reheat the universe either gravitationally \cite{Ford:1986sy,Feng:2002nb,BuenoSanchez:2007jxm,Matsuda:2007ax,Chun:2009yu,Dimopoulos:2018wfg} or through instant preheating (IP) \cite{Felder:1998vq,Campos:2002yk,Dimopoulos:2017tud}. These particles interact with the scalar field 
leading to SB at a temperature between $\sim 10^{8}-10^{12}$ GeV, which is higher than that of radiation era ($\sim 10^{7}$ GeV). In this course, the $\phi$-field spends longer time in the kination period than in inflation. Consequently, thermal equilibrium manifests before radiation fills the universe. The rate of baryon non-conserving process decreases, as the universe cools down due to Hubble expansion, which ceases to act at and below a certain temperature $T_F$, called the \textit{freeze-out temperature}. Therefore the temperature at which baryogenesis takes place, should be above $T_F$ (otherwise the CPT symmetry could be restored and baryon asymmetry would not occur). The actual value of $T_F$ \textit{i.e.} the energy scale of baryon non-conservation depends on the specific physical process involved, provided that the back reaction of the reheating particles on the scalar field dynamics is negligibly small \cite{Carroll:1998zi}, ensuring that the field velocity $\partial_0\phi$ does not vanish.\par In order to verify the quintessential SB scenario in the present and future experiments, the detection of relic gravitational waves (GWs) is the only choice. The GWs originate from the quantum fluctuations during inflation at sub-Hubble scale, which re-enter the Hubble horizon at different epochs. During transition from inflation to kination, due to a sharp change in the metric structure \cite{DeFelice:2002ir}, GWs of very high frequencies $\sim 10^{10}$ Hz, called \emph{blue-tilted} GWs \cite{Giovannini:1998bp,Giovannini:1999bh,Giovannini:1999qj,Riazuelo:2000fc,Figueroa:2018twl} are emitted, while during matter domination \emph{red-tilted} GWs of very low frequencies $\sim 10^{-17}-10^{-19}$ Hz are emitted \cite{Ahmad:2017itq,Ahmad:2019jbm}. The stochastic signatures of blue-tilted GWs are the evidences in favour of the quintessential inflationary model (specifically the kination period of expansion), while those of the GWs during radiation domination are instrumental in the verification of various BBN constraints \cite{Cyburt:2015mya,Figueroa:2018twl} \footnote{GWs having nano hertz frequencies are important in probing the origins of primordial blackhole, dark matter, compact binaries etc., which are supposed to be produced from secondary scalar perturbations in the form of induced gravitational waves. However, in that case the potential would have to be suitably modified in order to incorporate the ultra-slow-roll feature. Recently such approaches have been proposed in various literature. See Ref. \cite{Das:2023nmm} as an example.}. Overproduction of GWs would lead to spike in their energy densities \cite{Dimopoulos:2022wzo}, which could be threatening to nucleosynthesis process. These aspects specifically depend upon the particular modes of particle production involved after inflation. Specifically, gravitational reheating produces GW-energy too large to begin the nucleosynthesis, therefore inefficient to satisfy the required BBN constraints \cite{Cyburt:2015mya,Figueroa:2018twl}, whereas the IP is more efficacious in this regard. Another important aspect is the amplitudes of the GWs, called the \emph{characteristic strain}, which is extremely small $\sim 10^{-18}$ (see figure \ref{fig:Fig11}), thereby making the detection processes too difficult and complicated. In fact such a tiny value of the GW-amplitude carries indirect signatures of small values of tensor-to-scalar ratio and the amplitudes of CMB $B$ modes. These two quantities are again crucial for the conclusive proofs of the theory of inflation and associated phenomena. That is why, the detection of primordial GWs are the targets of many ongoing and forthcoming cosmological surveys. See sub-Section \ref{sec:Gravity_waves} for details. 
\par 
In an earlier work \cite{Sarkar:2023cpd}, we constructed a new model of quintessential $\alpha$-attractor, which has been shown to be efficacious in explaining the observational bounds of both inflation and dark energy (DE), simultaneously. The inflationary slow-roll regime has been probed by the analyses of the quasi de-Sitter quantum modes of the inflaton field by the `dynamical horizon exit method' and the corresponding mode responses of the cosmological parameters, like power spectra and spectral indices for a wide range of $\alpha$ from $\alpha=0.1$ to $\alpha=4.3$. The results for $k=0.002$ Mpc$^{-1}$ conform to the Planck-2018 data with $68\%$ CL. On the other hand, the calculated values of the vacuum density at the quintessential runaway are found to satisfy the requirement for explaining the present DE expansion of the universe, making the concerned model appropriate for slow-roll as well as quintessential inflation for the specified range of $\alpha$. The only portion which remains unexplored is the kination regime of the proposed model. Now, since this part of the potential is out of the inflationary regime, mode analysis is not useful. However, as discussed above, this section can act as the paradigm of SB \textit{vis-\`{a}-vis} the baryon asymmetry of the universe, if the scalar field derivatively interacts with a non-conserving baryon current in thermal equilibrium and the associated kinetic energy appears in the baryon excess as an extra contribution of particle density of the universe. Additionally, this part also plays the source of generating GW spectrum, which is an essential part of BBN during matter domination. This is in fact possible for necessary values of various temperatures and densities at different stages in the expansion history of the universe for initiating the SB process after inflation. They are required to obtain the experimentally favoured value of baryon-to-entropy ratio, which is a measure of baryon asymmetry of the present universe. A systematic exploration of the spontaneous baryogenesis scenario in kination regime can therefore unify three sections of the potential, \textit{viz.,} the inflationary slow-roll plateau, quintessential runaway and the steep kination regime in between them, as the resolutions of three monumental puzzles of standard model of cosmology - the origin of large scale structure, matter-antimatter asymmetry and the cosmological constant problem. Also, we can find the trend to what extent the parameter $\alpha$ will be constrained, when the three regimes will be present together within a single potential, considered here. Because precise determination of $\alpha$ has lots of connections with many other phenomena, among which fundamental origin of $\alpha$-attractor and specifically its interplay with the internal geometry of space-time is a rudimentary property of $\alpha$-attractor phenomenology. See Refs. \cite{Sarkar:2023cpd} and \cite{Sarkar:2021ird} for details. We shall also discuss on this aspect in Section \ref{sec:conclusion}.\par Motivated by these ideas, in the present paper we apply the model of quintessential $\alpha$-attractor, developed in Ref. \cite{Sarkar:2023cpd}, to the spontaneous baryogenesis scenario during kination by computing several parameters, within the range of $\alpha$ and model parameters indicated in same reference. Also, we study the amplitudes and frequencies of the relic GWs emitted at various epochs and their connections with the experimental constraints. Our aim will be to fix the $\alpha$ values which not only bring forth the necessary baryon-to-entropy ratio, but also be compatible with the results of the earlier study.\par The outline of the paper is as follows. In Section \ref{sec:kination}, we briefly review the kination period including the definitions of a number of temperatures in the context of post-inflationary particle production. Section \ref{sec:basic_framework} contains a concise description of SB. The new quintessential $\alpha$-attractor inflaton potential is applied to analyze the baryogenesis scenario, whose results are illustrated in Section \ref{sec:analysis}. The spectra of relic gravitational waves associated with this potential are also analyzed in the same section (sub-Section \ref{sec:Gravity_waves}). Finally, in Section \ref{sec:conclusion} we summarize the main results and write some concluding remarks.
 \section{The post-inflationary kination period\label{sec:kination}}
 In the standard formulation of inflationary quintessence \cite{Peebles:1998qn}, the $\phi$ field enters into a phase of kinetic energy domination, called \emph{kination} \cite{Joyce:1996cp}, just after the end of inflation. This period of expansion can be characterized by the scale factor ($a$)-dependent energy density \cite{Dimopoulos:2022wzo}
 \begin{equation}
     \rho_{\mathrm{kin}} (a)=\rho_{\mathrm{end}}\left(\frac{a_{\mathrm{end}}}{a}\right)^6
     \label{eq:kin_rho}
 \end{equation}
 and Hubble rate
 \begin{equation}
     H_{\mathrm{kin}}(a)=\sqrt{\frac{\rho_{\mathrm{kin}}(a)}{3M_p^2}}=H_{\mathrm{end}}\left(\frac{a_{\mathrm{end}}}{a}\right)^3.
     \label{eq:H_kin}
 \end{equation}
Here, the subscript `end' denotes the respective parameters at the point where the inflation ends and $M_p(=1/\sqrt{8\pi G})$ is the reduced Planck mass. The constants $\rho_{\mathrm{end}}$ and $H_{\mathrm{end}}$ are the inputs from the scale of inflation $V_{\mathrm{end}}$. Under the slow-roll approximation we can write
\begin{equation}
    \rho_{\mathrm{end}}\approx V_{\mathrm{end}},
    \label{eq:Correction1}
\end{equation}
because during inflation potential energy of the inflaton field is much larger than the kinetic energy. At the end of inflation, $\phi$ reaches $\phi_{\mathrm{end}}$ and the corresponding value of the potential, $V_{\mathrm{end}}=V(\phi_{\mathrm{end}})$, is obtained from equating the first potential slow-roll parameter $\epsilon_V$ to unity. The end-value of the potential is therefore induced in $\rho_{\mathrm{end}}$ and $H_{\mathrm{end}}$ as,
\begin{equation}  H_{\mathrm{end}}=\sqrt{\frac{\rho_{\mathrm{end}}}{3M_p^2}}\approx \sqrt{\frac{V_{\mathrm{end}}}{3M_p^2}}.
\label{eq:H_end}
\end{equation}\par Now, since in the period of kination, the kinetic energy of the $\phi$ field dominates over the potential energy (\emph{i.e.} $\frac{\Dot{\phi}^2}{2}\gg V(\phi)$), we can write
\begin{equation}
    \rho_{\mathrm{kin}}=\frac{\Dot{\phi}^2}{2}+V(\phi)\approx \frac{\Dot{\phi}^2}{2}.
    \label{eq:INT1}
\end{equation}
From Eqs. (\ref{eq:kin_rho}), (\ref{eq:Correction1}) and (\ref{eq:INT1}) we get
\begin{equation}
    \frac{\Dot{\phi}^2(a)}{2}\approx V_{\mathrm{end}}\left(\frac{a_{\mathrm{end}}}{a}\right)^6,
\end{equation}
which yields the field velocity of the form
\begin{equation}
    \Dot{\phi}(a)\approx \sqrt{2V_{\mathrm{end}}}\left(\frac{a_{\mathrm{end}}}{a}\right)^3.
    \label{eq:dot_phi}
\end{equation} 
Differentiating once again and identifying $H=\frac{\Dot{a}}{a}$ we obtain,
\begin{equation}
    \Ddot{\phi}+3H\Dot{\phi} = 0.
    \label{eq:Correction2}
\end{equation}
Eq. (\ref{eq:Correction2}) is the cosmological Klein-Gordon equation without the potential energy term. Therefore the dynamics of the $\phi$ field during kination is identical to that of a particle under free-fall, with the initial condition $\rho_{\mathrm{kin}}(a_{\mathrm{end}})=\rho_{\mathrm{end}}\approx V_{\mathrm{end}}$ or $\Dot{\phi}(a_{\mathrm{end}})\approx \sqrt{2V_{\mathrm{end}}}$ (see Eqs. (\ref{eq:kin_rho}), (\ref{eq:Correction1}) and (\ref{eq:dot_phi})), which is consistent with Eq. (\ref{eq:Correction1}) The dependency of $\Dot{\phi}$ on $a$, \textit{i.e.} $\Dot{\phi}\propto \frac{1}{a^3}$ is independent of the choice of inflaton potential. That is, the particular mathematical form or shape of the inflaton potential has no role during kination. From Eq. (\ref{eq:dot_phi}) it is also understandable that, the initial condition of the $\phi$-field dynamics and the magnitude of $\Dot{\phi}$, apart from the $a^{-3}$ dependency, vary with the scale of inflation $V_{\mathrm{end}}$. Generically $V_{\mathrm{end}}$ is a numerical constant and it is estimated by the COBE/Planck normalization condition (as we have also done in Ref. \cite{Sarkar:2023cpd}). Therefore, in general the kination scenario is model-independent. However, we shall see in Section \ref{sec:analysis} that specifically for quintessential $\alpha$-attractor potential, the $V_{\mathrm{end}}$ contains the free parameter $\alpha$, by which one can tune the magnitude of the value of $\Dot{\phi}$ along with almost all the other parameters of the kination period. This special feature of the said potential will be utilised to constrain the parameter $\alpha$ during kination. 
\par
Now, due to the expansion of the universe, $\Dot{\phi}$ becomes significantly small during kination with the increase in the scale factor $a$ and thereby the total energy density diminishes rapidly until radiation takes over through reheating associated with particle production. During the transition from inflation to kination, due to a non-adiabatic change of the metric structure \cite{DeFelice:2002ir}, an easy-to-realize mechanism of reheating is gravitational particle production (GPP) \cite{Ford:1986sy,Feng:2002nb,BuenoSanchez:2007jxm,Matsuda:2007ax,Chun:2009yu,Dimopoulos:2018wfg}. But it is not so effective so far as various constraints of BBN \cite{Dimopoulos:2017tud,Ahmad:2019jbm} are concerned. For example, GPP produces an insufficient amount of radiation density to start nucleosynthesis. GPP has been therefore replaced by an efficient method, called \emph{instant preheating} (IP) \cite{Felder:1998vq,Campos:2002yk,Dimopoulos:2017tud}. See Ref. \cite{Dimopoulos:2017tud} for details of this procedure, specifically for $\alpha$-attractor. Generally, it is an instantaneous (within one cycle of oscillation of the inflaton field), non-perturbative mechanism of reheating due to the effects of `\emph{broad parametric resonance}' (BPR) \cite{Kofman:1994rk,Kofman:1997yn,Greene:1997fu,Felder:1998vq}, characterized by fast energy transfer from the inflaton field to the superheavy massive particles produced, whose masses are typically $\sim 10^{17}-10^{18}$ GeV \cite{Felder:1998vq}. But in the case of quintessential inflation, as there is no minimum where the inflaton field can oscillate, rather a quintessential run-away, the BPR does not work. Just after the end of inflation, almost instantaneously the non-oscillating inflaton field produces superheavy particles, which again decay into fermions or bosons depending upon the nature of interactions. The elementary particles thus created give rise to a radiation density \cite{Sami:2004xk,Dimopoulos:2017tud,Ahmad:2019jbm,Dimopoulos:2022wzo} 
\begin{equation}  \rho_{\mathrm{rad}}(a)=\left[\frac{g M_p H_{\mathrm{end}}}{2\pi^{3/2}}\left(\frac{a_{\mathrm{end}}}{a}\right)^2\right]^2,
\label{eq:rad_density}
\end{equation} where $g(<1)$ is the positive coupling of IP. The associated temperature \cite{Kolb:1990vq,Dimopoulos:2022wzo} in unit of $M_p$ can be derived from Eq. (\ref{eq:rad_density}) as,
\begin{equation}  
    T(a)=\left(\frac{30}{\pi^2 g_{*}}\rho_{\mathrm{rad}}(a)\right)^{1/4}=\left(\frac{15g^2M_p^2H_{\mathrm{end}}^2}{2\pi^5g_{*}}\right)^{1/4}\left(\frac{a_{\mathrm{end}}}{a}\right).
  \label{eq:temp_density_2}
\end{equation}
Here, $g_{*}$ denotes the total number of degrees of freedom of the relativistic particles produced through IP. The factor in the first bracket containing constant terms has the dimension of temperature. Therefore it can be taken as a constant temperature term,
\begin{equation} T_{\mathrm{end}}=\left(\frac{15g^2M_p^2H_{\mathrm{end}}^2}{2\pi^5g_{*}}\right)^{1/4},
\label{eq:T_end}
\end{equation} such that,
\begin{equation}
    T=T_{\mathrm{end}}\left(\frac{a_{\mathrm
    end}}{a}\right).
    \label{eq:T_end_2}
\end{equation} $T_{\mathrm{end}}$ actually denotes the temperature at the instant of ending of inflation and beginning of kination at the scale factor $a=a_{\mathrm{end}}$. It depends on the energy scale of inflation, since, according to Eqs. (\ref{eq:H_end}) and (\ref{eq:T_end}) 
\begin{equation}
    T_{\mathrm{end}}=\left(\frac{5g^2V_{\mathrm{end}}}{2\pi^5 g_{*}}\right)^{1/4}.
\end{equation}\par Now, with the expansion of the universe, the decay rate with coupling $\beta$ \cite{Sami:2004xk,WaliHossain:2014usl,Ahmad:2019jbm} of the massive particles produced through IP decreases and let at $a=a_{\mathrm{th}}$ it becomes equal to the Hubble rate of expansion (see Eq. (\ref{eq:H_kin})). At this particular scale factor, thermal equilibrium commences and it is found to be,
\begin{equation}
    a_{\mathrm{th}}=\left(\frac{2^{6}\pi^6T_{\mathrm{end}}^4}{3^{3/2}g^3\beta^4M_p^3H_{\mathrm{end}}}\right)^{1/4}a_{\mathrm{end}},
\end{equation} which is, when applied in Eq. (\ref{eq:T_end_2}), gives the temperature at the thermal equilibrium,
\begin{equation}
T_{\mathrm{th}}=\left(\frac{3^{3/2}g^3\beta^4M_p^3H_{\mathrm{end}}}{2^{6}\pi^6}\right)^{1/4}.
\label{eq:thermal_T}
\end{equation}
This is the maximum temperature of the universe after inflation that can be attained in thermal equilibrium, called the \emph{maximum thermalization temperature} \cite{DeFelice:2002ir}. After this event, the total energy density of the scalar field becomes weaker and, as a consequence, radiation starts to emerge as a major component of the universe.\par Eq. (\ref{eq:T_end_2}) can be instrumental in evaluating the temperature during the era of radiation domination. Let us call the temperature as $T_{\mathrm{rad}}$ when the radiation domination starts and the corresponding scale factor as $a_{\mathrm{rad}}$. At this scale factor, there will be a \emph{cross-over} between two densities $\rho_{\mathrm{kin}}$ and $\rho_{\mathrm{rad}}$ (see Eqs. (\ref{eq:kin_rho}) and (\ref{eq:rad_density})), which means that at this stage $\rho_{\mathrm{kin}}(a_{\mathrm{rad}})=\rho_{\mathrm{rad}}(a_{\mathrm{rad}})$ (this happens at a particular value of e-folds) and thereafter $\rho_{\mathrm{rad}}$ will exceed $\rho_{\mathrm{kin}}$. Now, following the condition of cross-over from Eqs. (\ref{eq:kin_rho}) and (\ref{eq:rad_density}) we get
\begin{equation}
   a_{\mathrm{rad}}=\left(\frac{12\pi^3}{g^2}\right)^{1/2}a_{\mathrm{end}}
   \label{eq:EQ9}
\end{equation} and from Eqs. (\ref{eq:T_end_2}) and (\ref{eq:EQ9}) we have
\begin{equation}
    T_{\mathrm{rad}}=\left(\frac{g^2}{12\pi^3}\right)^{1/2}T_{\mathrm{end}}.
    \label{eq:T_rad_2}
\end{equation}
Eq. (\ref{eq:EQ9}) (Eq. (\ref{eq:T_rad_2})) shows that the ratio of the scale factors (temperatures) in the radiation domination era and the corresponding end-value depends only on the coupling constant of IP. In Section \ref{sec:analysis}, this finding will play an important role in analyzing the evolution of the universe during kination.\par The three temperatures $T_{\mathrm{end}}$, $T_{\mathrm{th}}$ and $T_{\mathrm{rad}}$ are crucial in jointly framing the whole scenario of the kination period. We can determine their values by doing a model-independent estimate. We take the help of an empirical formula \cite{Baumann:2009ds} relating the inflationary energy scale (in the unit of reduced Planck mass) with the tensor-to-scalar ratio $r$:
\begin{equation}
    V^{1/4}(r)\approx r^{1/4}\times \left(\frac{10^{16.5}}{2.43\times 10^{18}}\right)M_p.
    \label{eq:inf_scale}
\end{equation}
A good slow-roll inflaton potential always supports a small value of $r$ ($r<0.064$ \cite{Planck:2018vyg,Planck:2018jri}) and hence a small $\epsilon_V$, since $r=16\epsilon_V$. Therefore, it will be reasonable to consider a rough estimation $V\approx V_{\mathrm{end}}$, in a model-independent way. Thus, if we take $r=0.05$, then from Eqs.  (\ref{eq:inf_scale}) and (\ref{eq:H_end}), we obtain
\begin{equation}
    V_{\mathrm{end}}^{1/4}=6.15\times 10^{-3} M_p
\end{equation} and 
\begin{equation}
    H_{\mathrm{end}}=2.18\times 10^{-5} M_p.
\end{equation}
Choosing $\beta\sim 10^{-1}$, $g\sim 10^{-4}$ and $g_{*}\sim 10^2$ (based on various literature of IP, for example, \cite{Felder:1998vq,Dimopoulos:2017tud}), we finally obtain
\begin{equation}
    T_{\mathrm{end}}=5.84\times 10^{-6} M_p=1.42\times 10^{13}\quad \mathrm{GeV},
    \label{eq:EQ16}
\end{equation}
\begin{equation}
T_{\mathrm{th}}=6.55\times 10^{-7} M_p=1.59\times 10^{12}\quad \mathrm{GeV}
    \label{eq:T_therm}
\end{equation} and 
\begin{equation}
    T_{\mathrm{rad}}=3.32\times 10^{-11} M_p=8.06\times 10^{7}\quad \mathrm{GeV}.
    \label{eq:EQ18}
\end{equation}
 Eqs. (\ref{eq:EQ16}) - (\ref{eq:EQ18}) show that $T_{\mathrm{end}}/T_{\mathrm{th}}\approx 8.93$ and $T_{\mathrm{rad}}/T_{\mathrm{th}}\approx 5.07\times 10^{-5}$. Therefore it is clear that thermal equilibrium is reached shortly after the end of inflation and long before the advent of the radiation era, which is an essential requirement for the SB.
 \section{Basic framework of spontaneous baryogenesis}
 \label{sec:basic_framework}
The relevant Lagrangian here is,
 \begin{equation}
     \mathcal{L}_{\phi}=\left(-\frac{1}{2}g^{\mu\nu}\partial_{\mu}\phi\partial_{\nu}\phi - V(\phi) - \mathcal{L}_{\mathrm{eff}}\right),
     \label{eq:FirstEQ}
 \end{equation}
 where the quintessential inflaton field $\phi$ is in the Friedmann metric background $g_{\mu\nu}$ of signature $(-,+,+,+)$. In addition, $\phi$ is coupled with a non-conserving baryon current $J^\mu$ by the following $CP$-odd effective interaction \cite{DeFelice:2002ir,DeSimone:2016ofp,Ahmad:2019jbm,Basak:2021cgk}
 \begin{equation} \mathcal{L}_{\mathrm{eff}}=\frac{\lambda'}{M}\partial_{\mu}\phi J^{\mu},
 \label{eq:effective_int}
 \end{equation} $\lambda'$ being a dimensionless coupling constant and the mass scale $M<M_p$ signifies a sub-Planckian cut-off of the effective theory. The associated energy-momentum tensor can be obtained from the equation
 \begin{equation} T^{\mu}_{\nu}=\delta^{\mu}_{\nu}\mathcal{L}_{\phi}-2g^{\mu\alpha}\frac{\partial\mathcal{L}_{\phi}}{\partial g^{\alpha\nu}}.
     \label{eq:SecondEQ}
 \end{equation} \par The scalar field $\phi$ is spatially homogenized by the cosmic expansion after inflation. All primordial quantum fluctuations ($\delta\phi$) buried in $\phi$ within the Hubble sphere are transferred to density perturbations, whose imprints are detected in the anisotropies of the cosmic microwave background. The remaining $\phi$ field is now completely classical and time-dependent only. Therefore, in Eq. (\ref{eq:SecondEQ}) only $\mu=0$ component of $J^\mu$ \emph{viz.,} $J^0 = n-\Bar{n}$ will contribute, which measures the difference between the number densities of baryons ($n$) and ant-baryons ($\Bar{n}$). Following these arguments from Eq. (\ref{eq:SecondEQ}), the field density takes the form (see \ref{app:A})
 \begin{equation}
     \rho=-T_0^0=\left(\frac{\Dot{\phi}^2}{2}+V(\phi)\right)+\left(-\frac{\lambda'}{M}\Dot{\phi}\left(n-\Bar{n}\right)\right)
     =\rho_0+\left(n\delta\mathcal{E}+\Bar{n}\delta\Bar{\mathcal{E}}\right),
 \label{eq:baryo_density}
 \end{equation}
 where $\delta\mathcal{E}=-\delta\Bar{\mathcal{E}}=-\frac{\lambda'\Dot{\phi}}{M}$. $\rho_0$ is the interaction-free part of the density $\rho$. The second term indicates that, if the field velocity $\Dot{\phi}$ is appreciably non-zero, then for each baryon (anti-baryon), an extra amount of energy density $\delta\mathcal{E}(\delta\Bar{\mathcal{E}})$ would appear in the total energy density budget of the universe. The extra positive and negative energy density terms originate from the interaction term $\partial_\mu\phi J^\mu$ responsible for spontaneous $CPT$ symmetry breaking at a scale below the Planck mass, as explained in Section \ref{sec:intro}, which ultimately ensure the baryon-excess ($n>\Bar{n}$) in the universe. If, on the other hand, $CPT$ symmetry is not broken in thermal equilibrium, there would be no baryon asymmetry \emph{i.e.} $n=\Bar{n}$. We shall calculate $n-\Bar{n}$ in the next paragraph and show that it is positive definite. At the thermal equilibrium, this additional energy per baryon (anti-baryon) is recognized as an effective chemical potential $\mu$ ($-\mu$) \cite{DeSimone:2016ofp}. Alternative realizations of chemical potentials, in this context, can be seen in Refs. \cite{Arbuzova:2016qfh,Dasgupta:2018eha}. \par We explicitly compute the baryon excess by considering the statistical phase space distribution function in thermal equilibrium,
  \begin{equation}
     f (\epsilon,\mu)=\frac{1}{e^{\left(\frac{\epsilon-\mu}{T}\right)}+1}=\frac{1}{e^{\left(\frac{|\Vec{p}|-\mu}{T}\right)}+1},   
 \end{equation}
  of an ensemble of massless relativistic baryons of momentum $\Vec{p}$ at a particular energy level $\epsilon=|\Vec{p}|$ with degeneracy $\Tilde{g}$ as,
 \begin{equation}
     \Delta n = n-\Bar{n}=\int \Tilde{g}\frac{d^3\Vec{p}}{(2\pi)^3}\left[f(\epsilon,\mu)-f(\epsilon,-\mu)\right],
     \label{eq:baryon_excess}
 \end{equation}
 with the convention $c=\hbar = k_B =1$. Under the high temperature approximation, $\mu/T\ll 1$, Eq. (\ref{eq:baryon_excess}) reduces to (see \ref{app:B})
 \begin{equation}
     \Delta n \approx \frac{\Tilde{g}\mu T^2}{6}=\frac{\Tilde{g}\lambda'\Dot{\phi}T^2}{6M}>0.
     \label{eq:B_excess}
 \end{equation}
 The temporal behavior of $\phi$, determined by $\Dot{\phi}$, is encoded in the solution of its equation of motion under the effective interaction of Eq. (\ref{eq:effective_int}) during kination in expanding background. We derive that equation of motion, using Eq. (\ref{eq:FirstEQ}) from the Euler-Lagrange equation
 \begin{equation}
   \left[\partial_{\mu}\left(\frac{\partial}{\partial_\mu\phi}\right) -\frac{\partial}{\partial\phi}\right]\sqrt{-g}\mathcal{L}_\phi=0,
 \end{equation}
 as (see \ref{app:C})
 \begin{equation}
\left[1-\frac{\Tilde{g}}{6}\left(\frac{\lambda'M_p}{M}\right)^2\left(\frac{T}{M_p}\right)^2\right]\left(\Ddot{\phi}+3H\Dot{\phi}\right)+\frac{d V(\phi)}{d\phi}=0.
     \label{eq:phi_eqn_1}
 \end{equation}
 The second term in the third bracket signifies the back-reaction on $\phi$ of the coupling described in Eq (\ref{eq:effective_int}). According to the \emph{Carroll bound} \cite{Carroll:1998zi,Trodden:2003yn} $\frac{\lambda'M_p}{M}<8$, and as described in Section \ref{sec:intro}, the baryon asymmetry operates at temperature $T$ lying in the range $T_F<T<T_{\mathrm{th}}$, where $T_{\mathrm{th}}\sim 10^{-7} M_p$ (see Eq. (\ref{eq:T_therm})), which implies $T/M_p\ll 1$. Therefore the effect of back-reaction \emph{vis-\`{a}-vis} the second term is very small. Then from Eq. (\ref{eq:phi_eqn_1}) the dynamical equation of $\phi$ reduces to
 \begin{equation}
     \Ddot{\phi}+3H\Dot{\phi}+\frac{dV(\phi)}{d\phi}\approx0,
     \label{eq:EQ31}
 \end{equation}
which is the well-known Klein-Gordon equation of the scalar field $\phi$. \par Now, in the kination regime, $\frac{\Dot{\phi}^2}{2}\gg V(\phi)$. Therefore the potential gradient $\frac{dV(\phi)}{d\phi}$ is negligible compared to $\Dot{\phi}$ and $\Ddot{\phi}$. Thus we get,
 \begin{equation}
\Ddot{\phi}+3H\Dot{\phi}\approx0.
\label{eq:DayanaPhi}
 \end{equation} 
Considering $H=\frac{\Dot{a}}{a}$ in Eq. (\ref{eq:DayanaPhi}) we obtain,
\begin{equation}
\frac{1}{a^3}\frac{d}{dt}\left(a^3\Dot{\phi}\right)=0,
\end{equation}
whose solution is given by
\begin{equation}
    \Dot{\phi}(a)=\frac{C}{a^3},
    \label{eq:EQ32}
\end{equation}
 where $C$ is a constant. Applying $\Dot{\phi}_{\mathrm{end}}=\Dot{\phi}(a_{\mathrm{end}})\approx \sqrt{2V_{\mathrm{end}}}$ as initial condition, we obtain $C=\sqrt{2V_{\mathrm{end}}}\left(a_{\mathrm{end}}\right)^3$ and thus Eq. (\ref{eq:EQ32}) matches with Eq. (\ref{eq:dot_phi}). Therefore, under negligible amount of back-reaction, the effective interaction gives the same equation of kination, as described in Section \ref{sec:kination}. This indicates that the framework of SB is fairly consistent with the scenario of kination.\par According to Eqs. (\ref{eq:dot_phi}) and (\ref{eq:T_end_2}) the field velocity $\Dot{\phi}$ can be parameterized as a function of temperature as
 \begin{equation}
     \Dot{\phi}(T)=\sqrt{2V_{\mathrm{end}}}\left(\frac{T}{T_{\mathrm{end}}}\right)^3.
     \label{eq:T_phidot}
 \end{equation} Eqs. (\ref{eq:B_excess}) and (\ref{eq:T_phidot}) show that the baryon excess $\Delta n$ is only temperature dependent. It is conveniently measured with respect to the entropy density $s$ (given in Refs. \cite{DeSimone:2016ofp,Ahmad:2019jbm,Basak:2021cgk}) as
 \begin{equation}
     \eta_T =\left(\frac{\Delta n}{s}\right)=\frac{\left(\frac{\Tilde{g}\lambda'\Dot{\phi}T^2}{6M}\right)}{\left(\frac{2\pi^2g_{*}T^3}{45}\right)}
     =\frac{15}{4\pi^2}\left(\frac{\lambda'}{M}\right)\left(\frac{\Tilde{g}}{g_*}\right)\left(\frac{\Dot{\phi}(T)}{T}\right).
     \label{eq:BTER_1}
 \end{equation} 
 This is called the \textit{baryon-to-entropy ratio} (BTER) at a finite temperature $T$. Since $\Dot{\phi}(T)\propto T^3$, $\eta_T\propto T^2$, \emph{i.e.} $\eta_T$ decreases as the temperature of the universe decreases with expansion. But, as explained in Section \ref{sec:intro}, this will continue up to the freeze-out temperature $T_F$. Below $T_F$ the baryon non-conserving processes fall out of thermal equilibrium and no further baryons are generated. As a result, $\eta_F$ saturates at a freeze out value
 \begin{equation}
     \eta_{F} =\frac{15}{4\pi^2}\left(\frac{\lambda'}{M}\right)\left(\frac{\Tilde{g}}{g_*}\right)\left(\frac{\Dot{\phi}(T_F)}{T_F}\right),
     \label{eq:BTER_2}
 \end{equation}
 which acts as a standard measure of the observed baryon asymmetry in the present universe. Therefore it should conform to the experimental value of $\eta$, mentioned in Section \ref{sec:intro}. \par Following Eqs. (\ref{eq:H_end}), (\ref{eq:T_end}), (\ref{eq:T_end_2}), (\ref{eq:thermal_T}) and (\ref{eq:T_phidot}) the freeze out value of the field velocity takes the form
 \begin{equation}
       \Dot{\phi}(T_F)=\sqrt{6M_p^2}H_{\mathrm{end}}\left(\frac{T_F}{T_{\mathrm{end}}}\right)^3
     =\left(\frac{3}{2}\right)^{5/4}2M_p\beta^2\left(\frac{\pi g_{*}}{60}\right)^{3/4}\frac{T_F^3}{T_{\mathrm{th}}^2}.
 \label{eq:phi_dot_T}
 \end{equation}
 We now finally obtain from Eqs. (\ref{eq:BTER_2}) and (\ref{eq:phi_dot_T}),
 \begin{equation}
      \eta_F=\left(\frac{3}{2}\right)^{5/4}\left( \frac{30}{4\pi^2}\right)\left(\frac{\pi g_{*}}{60}\right)^{3/4}\left(\frac{\Tilde{g}}{g_{*}}\right)\left(\frac{\lambda'M_p}{M}\right)\left(\frac{\beta T_F}{T_{\mathrm{th}}}\right)^2
      =4.36635\times 10^{-4}\times \left(\frac{\lambda'M_p}{M}\right)\left(\frac{T_F}{T_{\mathrm{th}}}\right)^2,
 \label{eq:BTER_final}
 \end{equation}
 where $\Tilde{g}\sim 1$ and we put the values of $g_{*}$ and $\beta$, used for evaluating the temperatures of Eqs. (\ref{eq:EQ16}) - (\ref{eq:EQ18}). Here, $T_{\mathrm{th}}$ bears the signature of the scale of inflation $V_{\mathrm{end}}$, while $M$ and $T_F$ are specific to the particular baryon non-conserving process involved. Thus, in precise evaluation of $\eta_F$ in light of the observed value, the data for both the inflationary energy scale and the baryon asymmetry phenomenology play the determining roles. In order to keep the thermal equilibrium during baryogenesis, the $T_F$ should lie within $T_{\mathrm{rad}}$ and $T_{\mathrm{th}}$. In fact, by the Carroll bound, a rough estimate of $T_F$ can be computed as $T_F=1.03\times 10^{-10} M_p=2.5\times 10^{8}$ GeV, if $T_{\mathrm{th}}$ and $\eta_F$ are allowed to take the model-independent value (see Eq. \ref{eq:T_therm}) and the observational bound (mentioned in Section \ref{sec:intro}), respectively.\par The end-value of the potential $V_{\mathrm{end}}$ acts as the initial condition for kination and SB. In most of the cases, $V_{\mathrm{end}}$ is a fixed quantity, determined by COBE/Planck normalization. Therefore, the process of SB is model independent. However, as will be shown in next section, in case of quintessential $\alpha$-attractor or specifically for a particular form of quintessential $\alpha$-attractor \cite{Sarkar:2023cpd}, $V_{\mathrm{end}}$ is $\alpha$-dependent. As a result, an extra provision is generated through $V_{\mathrm{end}}$ to control the magnitude of BTER. In fact, almost all the parameters of kination and SB described in Section \ref{sec:kination} and in the present section, become functions of $\alpha$ though $V_{\mathrm{end}}$ for the model indicated earlier. Thus varying $\alpha$, one can tune those parameters to obtain the required experimental value of BTER. This in turn constrain $\alpha$ in light of experimental perspectives, which is actually our aim in studying the role of $\alpha$ in SB during kination. Therefore, in next section we shall apply the end-value expression of a specific model of quintessential $\alpha$-attractor and study the kination and SB scenarios.
 \section{Results and discussion}
 \label{sec:analysis}
Let us begin with the following quintessential $\alpha$-attractor potential, developed recently in Ref. \cite{Sarkar:2023cpd}
\begin{equation}
    V(\phi)=V_{*}(n,\alpha)e^{-n\left(1-\tanh{\frac{\phi}{\sqrt{6\alpha}M_p}}\right)}.
    \label{eq:Qpot}
\end{equation}
 $V_{*}(n,\alpha)$ is the COBE/Planck-normalized inflationary energy scale, which is a function of model parameters $n$ and $\alpha$ and has been derived in Ref. \cite{Sarkar:2023cpd}. The associated first potential slow-roll parameter is found to be
\begin{equation}
    \epsilon_V(\phi)=\left[\left(\frac{n}{\sqrt{12\alpha}}\right)^{1/2}\operatorname{sech}{\frac{\phi}{\sqrt{6\alpha}M_p}}\right]^{4}.
\end{equation}
When $\phi=\phi_{\mathrm{end}}$, $\epsilon_V(\phi_{\mathrm{end}})=1$, from which we obtain the end-value of $\phi$:
\begin{equation}
    \phi_{\mathrm{end}}=\sqrt{6\alpha}M_p\operatornamewithlimits{sech^{-1}}\left(\frac{\sqrt{12\alpha}}{n}\right)^{1/2}
\end{equation}
and the corresponding end-value expression of the potential of Eq.(\ref{eq:Qpot}) is
\begin{equation}
    V_{\mathrm{end}}(n,\alpha)=V_{*}(n,\alpha)e^{-n\left[1-\sqrt{1-\left(\frac{\sqrt{12\alpha}}{n}\right)}\right]}.
    \label{eq:Vend}
\end{equation}
\begin{figure}[H]
	\centering
\includegraphics[width=0.8\linewidth]{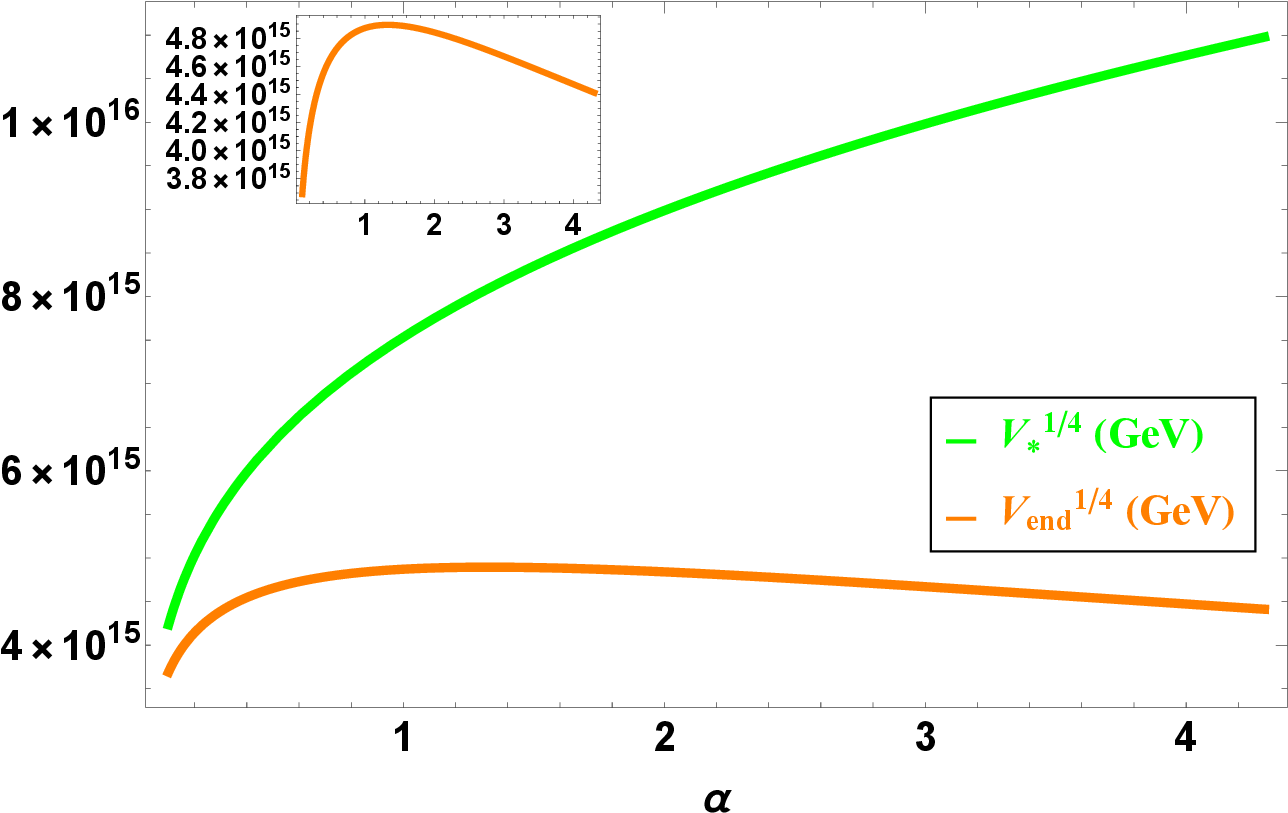}
	\caption{The beginning and the end of the energy scales of inflation in GeV for $n=122$, plotted against $\alpha$ between $0.1$ and $4.3$. The small variation of $V_{\mathrm{end}}^{1/4}$ near the value $4\times 10^{15}$ GeV is shown in the inset.}
	\label{fig:Fig1}
\end{figure}
Figure \ref{fig:Fig1} shows the variations of $V_{*}^{1/4}$ and $V_{\mathrm{end}}^{1/4}$ with $\alpha$ from $\alpha=0.1$ to $4.3$ for $n=122$. The choices of $\alpha$ and $n$ are based upon the results in Ref. \cite{Sarkar:2023cpd}. Depending upon the values of $\alpha$, inflation occurs at a scale $V_{*}^{1/4}$, lying between $\sim10^{15}-10^{16}$ GeV (see the green curve), but it ends at a scale $V_{\mathrm{end}}^{1/4}$, which is strictly around $\sim10^{15}$ GeV (see the orange curve) for all allowed values of $\alpha$, which is also the point of initialization of the kination period. As $\alpha$ increases, the difference between the two energy scales \emph{viz.,} $V_{*}^{1/4}$ and $V_{\mathrm{end}}^{1/4}$ also increases because of the \emph{double pole behavior} (discussed in Ref. \cite{Sarkar:2023cpd}) of the inflaton potential of Eq. (\ref{eq:Qpot}). When $\alpha$ is small, the potential is of slow-roll type, where the beginning and the end of energy scales of inflation are almost of the same order. But as $\alpha$ increases, the potential becomes of the power-law type and the two energy scales start to differ up to $\sim 10$ GeV. The order of magnitude of $V_{\mathrm{end}}^{1/4}$ is instrumental in obtaining correct ranges of the energy densities and temperatures of the universe during kination, as will be described in next subsections. In the subsequent works we shall fix the $n$ value to $n=122$ and vary the $\alpha$ parameter in the range $0.1\leq\alpha\leq 4.3$. Because we wish to study the SB process during kination within the range of $\alpha$, where we obtained good results of early-time and late-time expansions of the universe in Ref. \cite{Sarkar:2023cpd}.
\subsection{Evolution of the energy densities}
The evolution of various energy densities in the kination period primarily depends upon the ratio $a/a_{\mathrm{end}}$, which is measured through the number of e-folds $N$, elapsed after the end of inflation,
\begin{equation}
    N\equiv\int_{0}^{N}dN'=\int_{a_{\mathrm{end}}}^{a}d\ln{a'}=\ln{\left(\frac{a}{a_{\mathrm{end}}}\right)}.
    \label{eq:e_folds}
\end{equation}
Now, from Eqs. (\ref{eq:kin_rho}), (\ref{eq:H_kin}), (\ref{eq:H_end}), (\ref{eq:rad_density}) and (\ref{eq:e_folds}) we get
\begin{equation}
    \left(\frac{\rho_{\mathrm{kin}}}{V_{\mathrm{end}}}\right)=e^{-6N},
\end{equation}
\begin{equation}
    \left(\frac{H_{\mathrm{kin}}^2M_p^2}{V_{\mathrm{end}}}\right)=\frac{1}{3}e^{-6N}
\end{equation} and 
\begin{equation}
    \left(\frac{\rho_{\mathrm{rad}}}{V_{\mathrm{end}}}\right)=\frac{g^2}{12\pi^3}e^{-4N}.
\end{equation}
\begin{figure}[H]
	\centering
\includegraphics[width=0.8\linewidth]{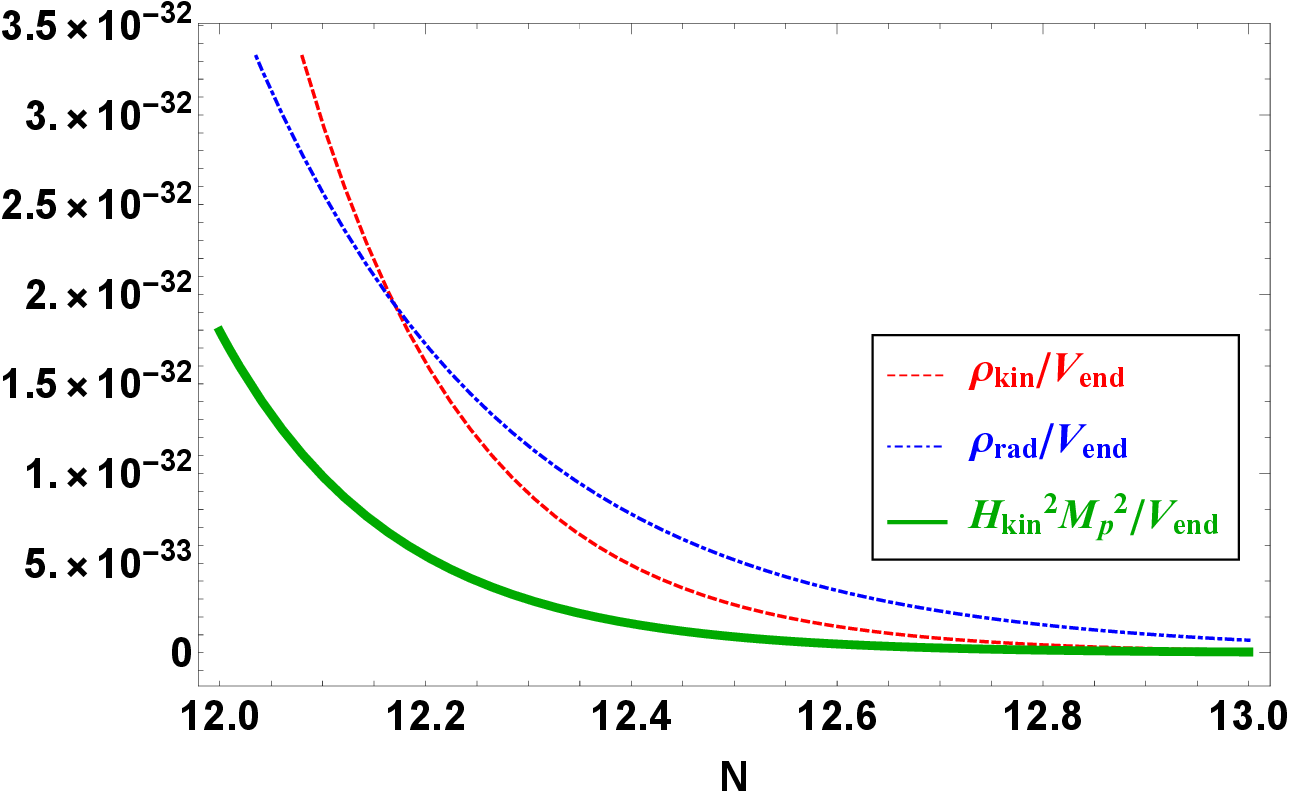}
	\caption{Evolution of the dimensionless post-inflationary energy densities with the number of e-folds. At $N\approx 12.17$, the total energy density of the quintessential inflaton field and that of radiation become equal and after that radiation starts to dominate \emph{i.e.} radiation era begins.}
	\label{fig:Fig2}
\end{figure}
In figure \ref{fig:Fig2} the behaviors of the energy density functions, stated above, are described with the expansion of the universe, measured by the number of e-folds $N$. It is clear that, $\rho_{\mathrm{kin}}\sim\rho_{\mathrm{rad}}=10^{-32}V_{\mathrm{end}}\sim 10^{28}$ (Gev)$^{4}$. This amount of energy density is sufficient for satisfying the observational data of primordial nucleosynthesis \cite{DeFelice:2002ir,Basak:2021cgk}, which has become possible for selecting the IP as a proper reheating mechanism instead of the gravitational reheating or GPP. Such a choice therefore sets a perfect stage for the successful SB process, which will be understood more clearly later. Another interesting observation in figure \ref{fig:Fig2} is that, there is a cross-over between $\rho_{\mathrm{kin}}$ and $\rho_{\mathrm{rad}}$ at $N\approx 12.17$. The radiation density begins to be greater than the field density when the expansion of the universe reaches $12.17$ numbers of e-folds after inflation ends. Even more crucial is that, this value is model-independent (which will be explained in the next sub-section) \emph{i.e.} for all models, the radiation domination occurs after $12.17$ e-folds from the end of inflation. Only the magnitudes of the said energy densities at that specific number of e-folds depend upon the scale of the inflaton potential concerned. This value of $N$ limits the maximum amount of expansion, below which the SB takes place. Once the radiation density becomes larger than the field density, the process of baryogenesis stops. To get acquainted with this scenario with further clarity, the evolution of temperatures of the universe at different phases should be carefully examined.
\subsection{Evolution of the temperatures}
\label{subsec:temperatures}
\begin{figure}[H]
    \begin{subfigure}{0.5\linewidth}
     \centering
\includegraphics[width=70mm,height=50mm]{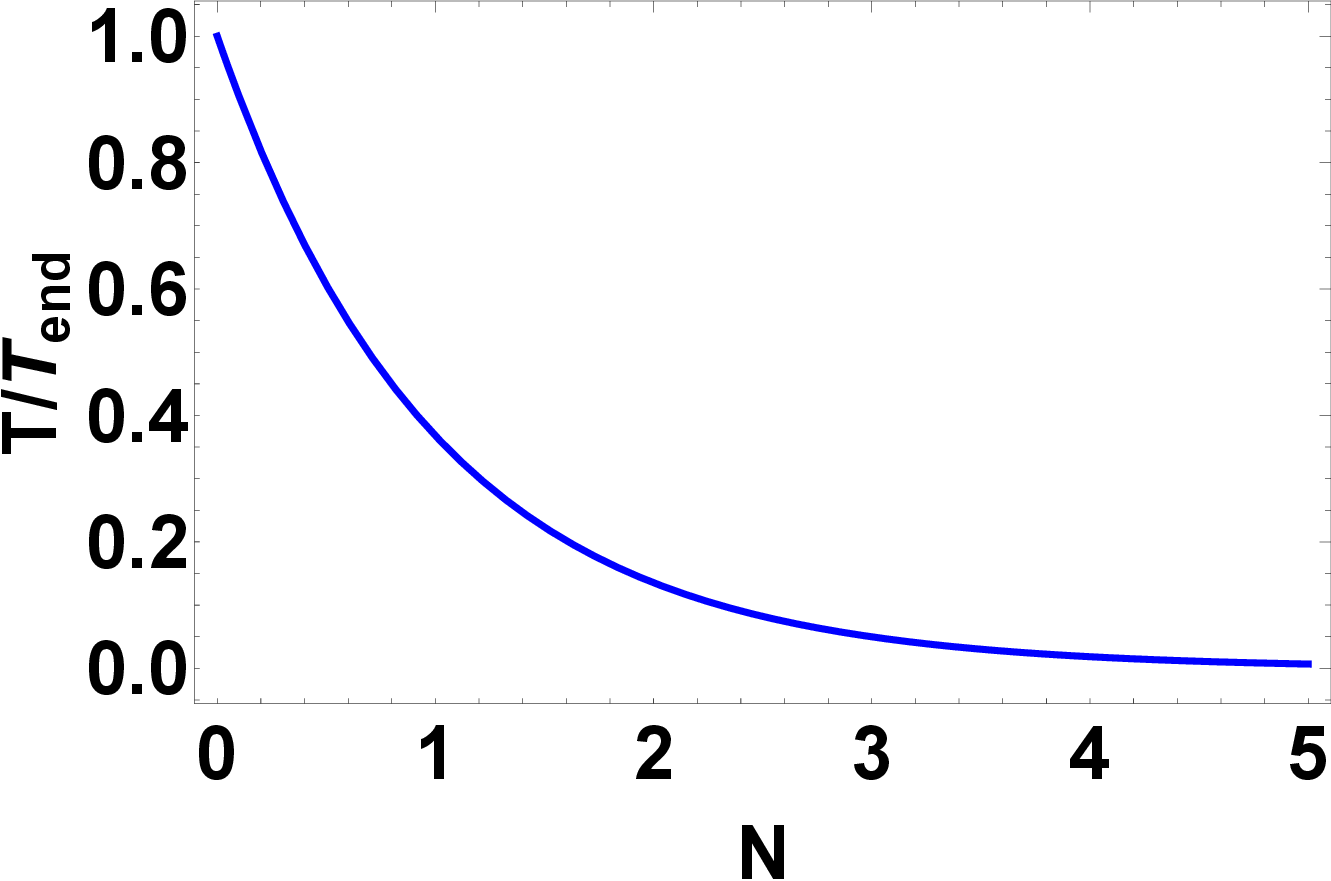}
  \subcaption{}
   \label{fig:Fig3a}
\end{subfigure}
\begin{subfigure}{0.5\linewidth}
\centering
\includegraphics[width=70mm,height=50mm]{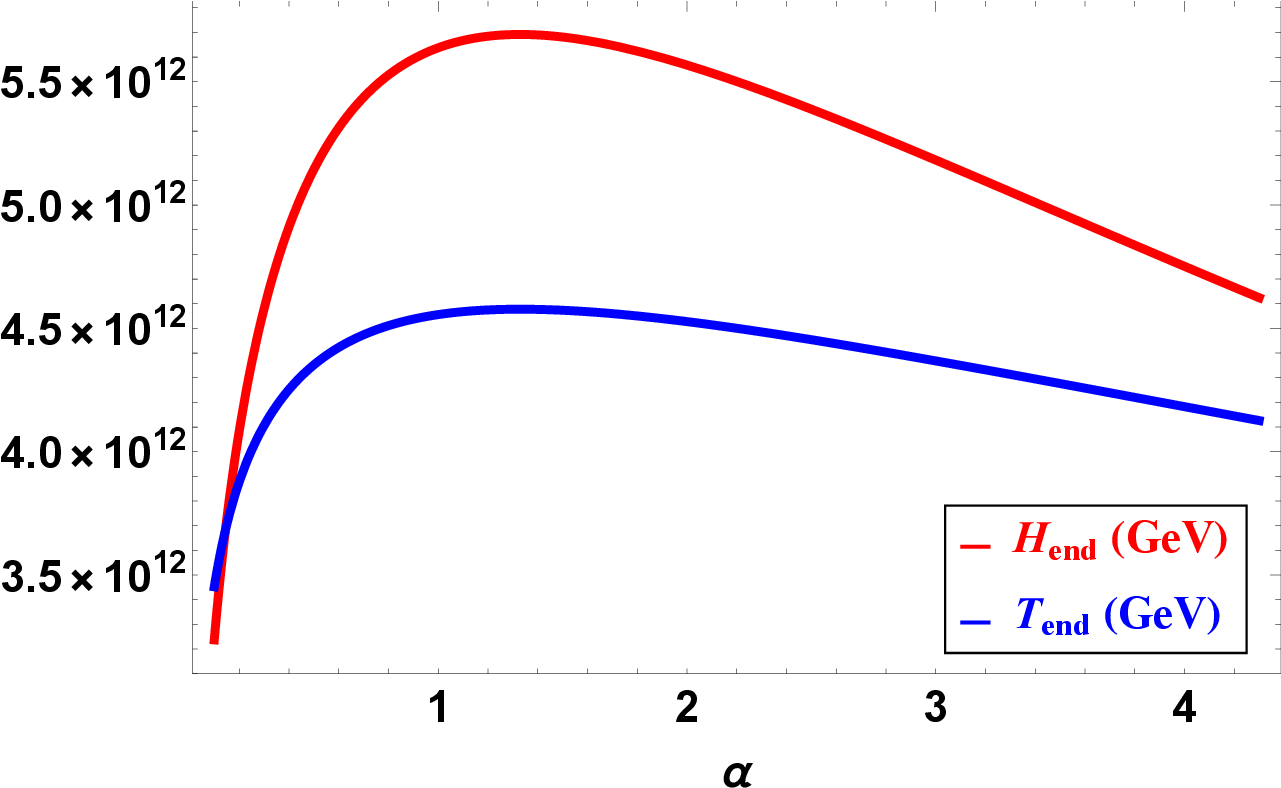}
   \subcaption{}
    \label{fig:Fig3b}
\end{subfigure}
     \caption{\textit{Left panel}: Variation of temperature with number of e-folds $N$ starting from $T=T_{\mathrm{end}}$. \textit{Right panel}: The $\alpha$-dependencies of $T_{\mathrm{end}}$ and $H_{\mathrm{end}}$ indicate that kination starts at temperature $\sim 10^{12}$ GeV. Around $\alpha=1$, the values of $H_{\mathrm{end}}$ and $T_{\mathrm{end}}$ become maximum.}
    \label{fig:Fig3}
\end{figure}

In the left panel of figure \ref{fig:Fig3}, the evolution of $T/T_{\mathrm{end}}$ is shown as a function of number of e-folds $N$, which is independent of the values of $\alpha$. According to Eqs. (\ref{eq:T_end_2}) and (\ref{eq:e_folds}) the temperature of the universe falls exponentially after the completion of inflation, from $T=T_{\mathrm{end}}$ (see Eq. (\ref{eq:T_end})). In the right panel of figure \ref{fig:Fig3}, the dependencies of $T_{\mathrm{end}}$ and $H_{\mathrm{end}}$ on $\alpha$ are shown. $T_{\mathrm{end}}$ is $\sim 10^{12}$ GeV for all values of $\alpha$ and $H_{\mathrm{
end}}$ denotes the expansion rate at the end of inflation. This is the temperature at the entry-level (since here $N=0$) of the kination period.
Now, as the universe enters into kination, thermal equilibrium is not attained soon. It requires some time (or some extra e-folds), which is needed for the initiation of the interaction of the field with the particles produced via reheating. The maximum thermalization temperature $T_{\mathrm{th}}$ (see Eq. (\ref{eq:thermal_T})) determines the threshold of this equilibrium. 
\begin{figure}[H]
    \begin{subfigure}{0.5\linewidth}
  \centering
\includegraphics[width=70mm,height=50mm]{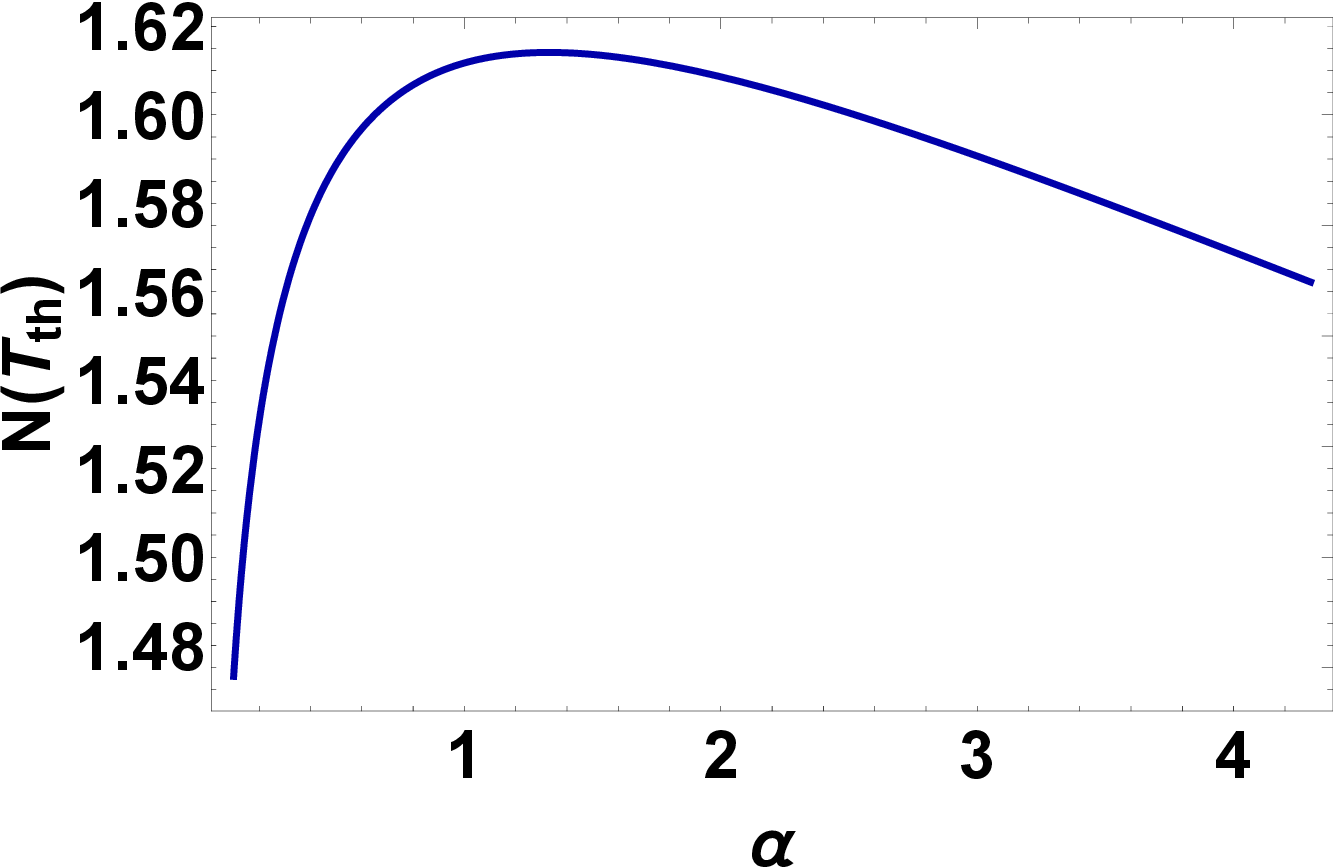}
  \subcaption{}
   \label{fig:Fig4a}
\end{subfigure}
\begin{subfigure}{0.5\linewidth}
  \centering
\includegraphics[width=70mm,height=50mm]{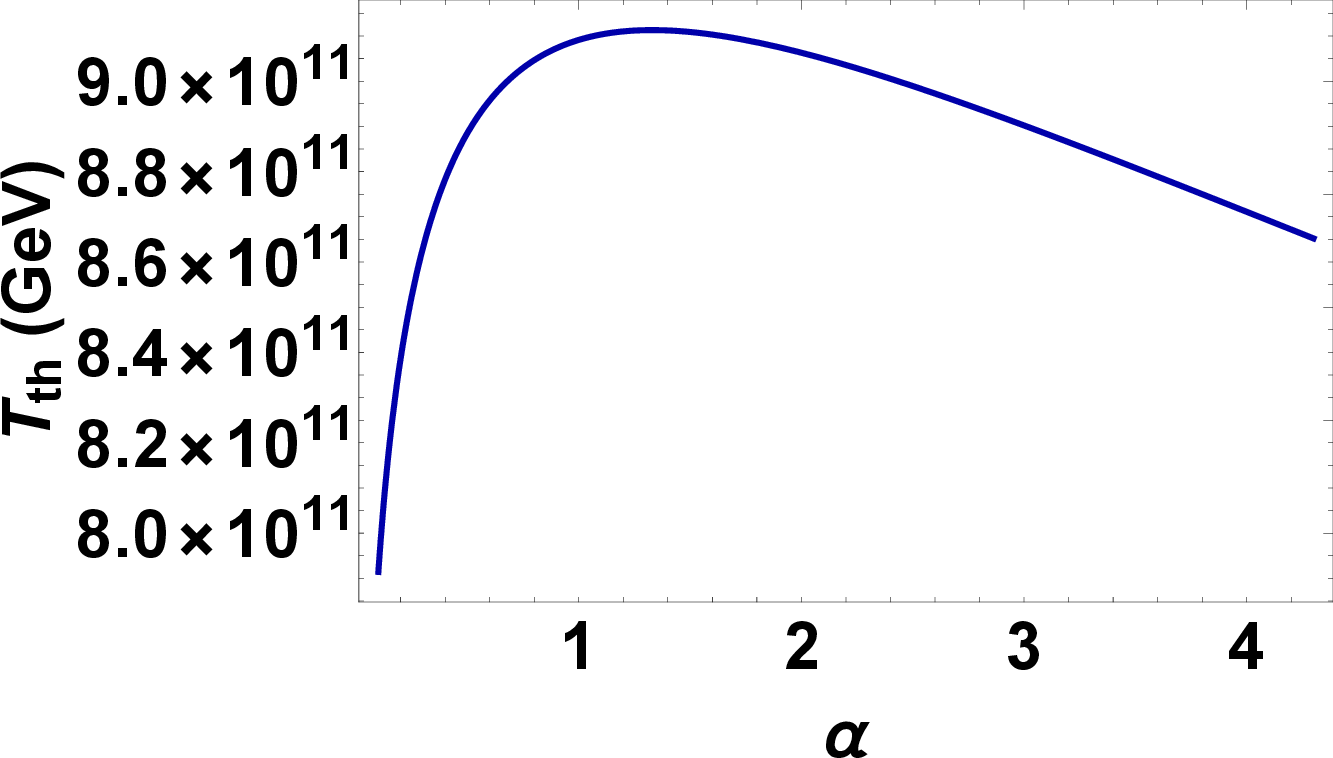}
   \subcaption{}
    \label{fig:Fig4b}
\end{subfigure}
     \caption{\emph{Left panel}: The number of e-folds required to attain the thermal equilibrium at $T=T_{\mathrm{th}}$. \emph{Right panel}: The dependency of $T_{\mathrm{th}}$ on $\alpha$ depicts the fact that thermal equilibrium starts at a temperature around $10^{11}$ GeV.}
    \label{fig:Fig4}
\end{figure}
If the $N(T_{\mathrm{th}})$ is the number of e-folds passed to reach the thermal equilibrium, then from Eq. (\ref{eq:T_end_2}) and (\ref{eq:e_folds}),
\begin{equation}
    N(T_{\mathrm{th}})=\ln{\left(\frac{T_{\mathrm{end}}}{T_{\mathrm{th}}}\right)},
\end{equation}
which is roughly $1.47-1.61$ depending on $\alpha$ chosen for the potential and the associated maximum thermalization temperature is $\sim 10^{11}$ GeV (see figure \ref{fig:Fig4}). Therefore, approximately after $2$ e-folds, when the temperature drops from $10^{12}$ GeV to $10^{11}$ GeV, equilibrium is achieved. After that, the quintessential baryogenesis takes place by the
spontaneous $CPT$ symmetry breaking, as explained in Sections \ref{sec:intro} and \ref{sec:basic_framework}. Therefore, $T_{\mathrm{th}}$ refers to the maximum possible temperature below which baryogenesis takes place.\par The baryon non-conserving process remains active until the radiation does not overpower the total field density and thereafter, when the radiation density crosses the field density at temperature $T_{\mathrm{rad}}$ (see Eq. (\ref{eq:T_rad_2})), the process of baryogenesis stops. Thus, SB should occur above $T_{\mathrm{rad}}$ and it therefore acts as the minimum possible temperature above which baryogenesis must occur. Actually, the baryon number violating process ceases to act, specifically at and below the freeze-out temperature $T_F$, which will be shown in next sub-section lying between $T_{\mathrm{th}}$ and $T_{\mathrm{rad}}$.\par  
\begin{figure}[H]
    \begin{subfigure}{0.5\linewidth}
  \centering
\includegraphics[width=70mm,height=50mm]{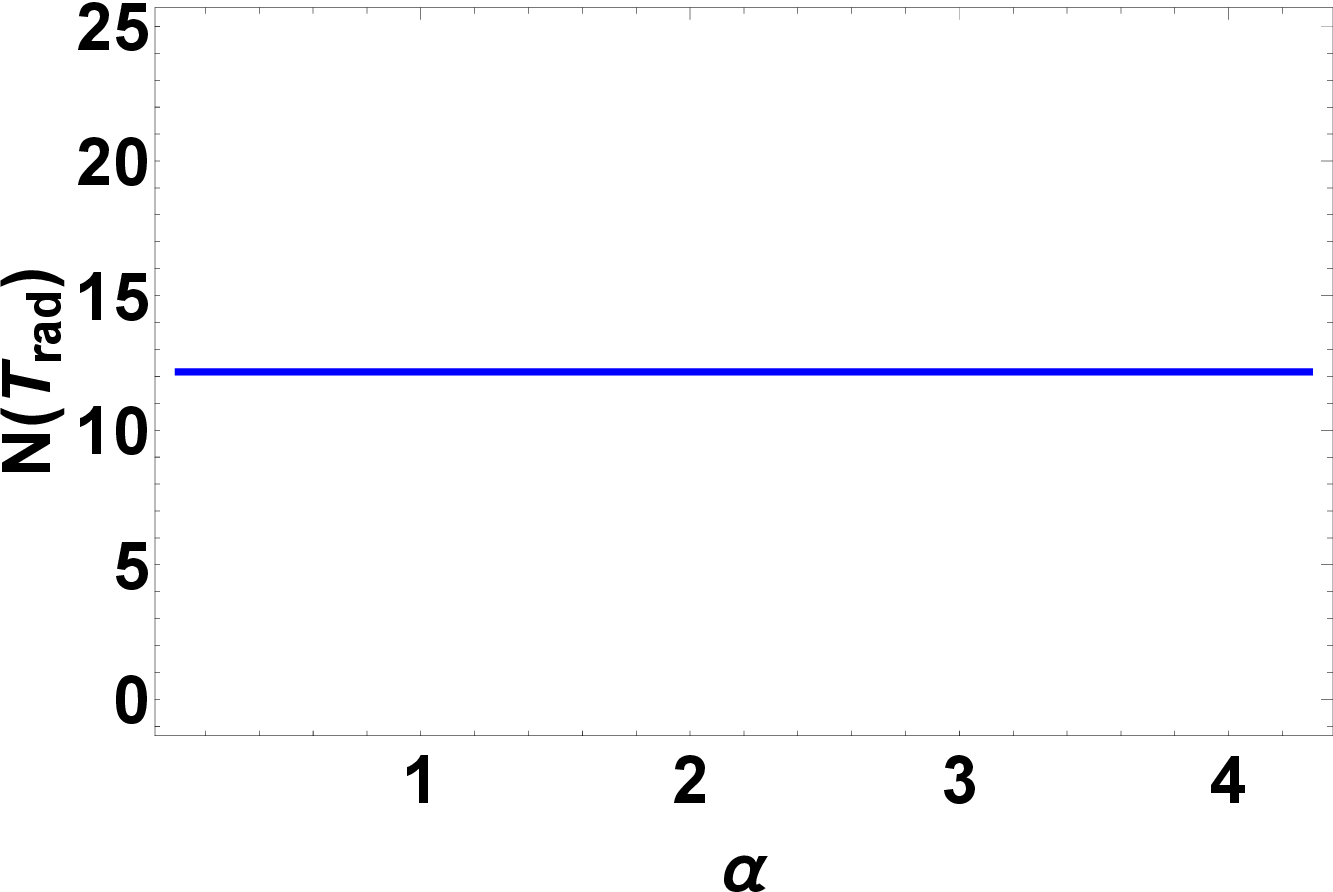}
  \subcaption{}
   \label{fig:Fig5a}
\end{subfigure}
\begin{subfigure}{0.5\linewidth}
  \centering
\includegraphics[width=70mm,height=50mm]{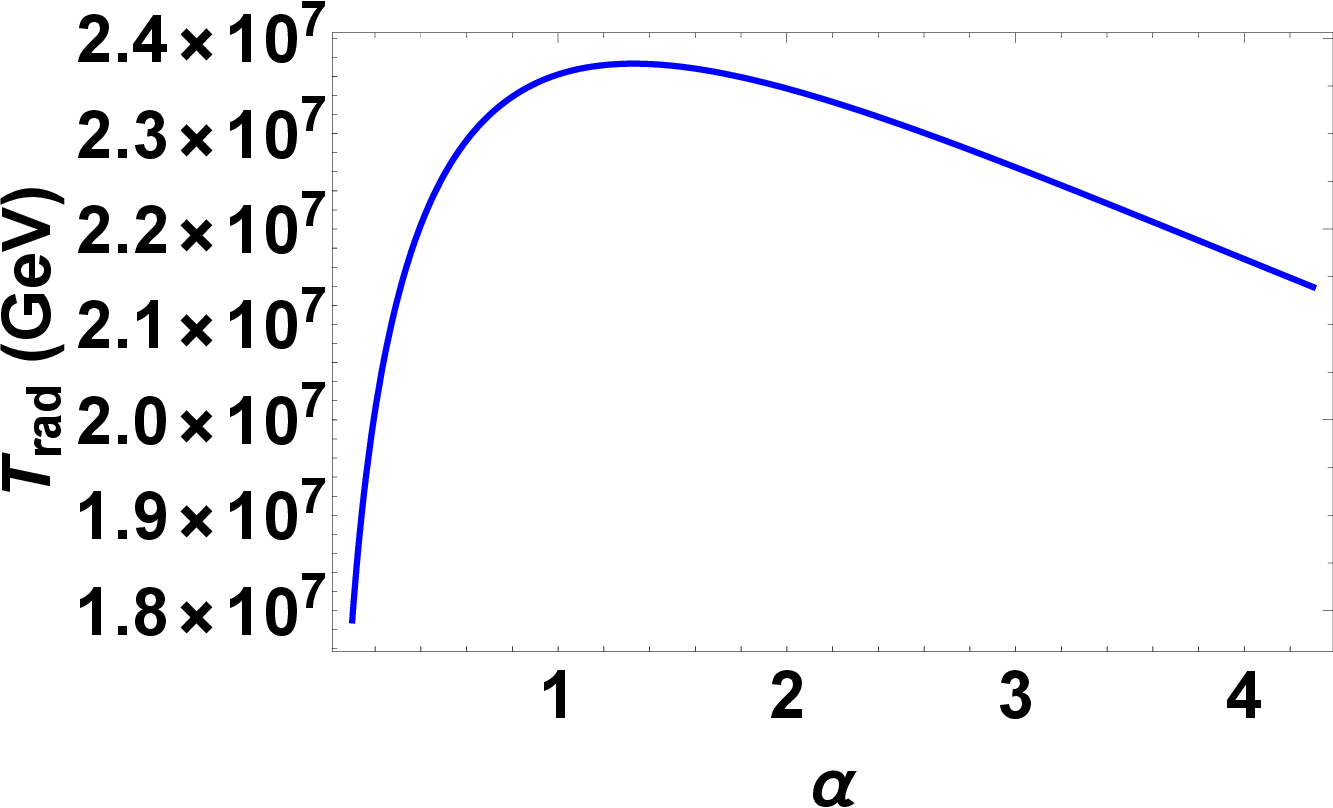}
   \subcaption{}
    \label{fig:Fig5b}
\end{subfigure}
     \caption{\emph{Left panel}: The number of e-folds needed to begin the radiation domination at $T=T_{\mathrm{rad}}$.  \emph{Right panel}: The dependency of $T_{\mathrm{rad}}$ on $\alpha$ depicts that the radiation domination is found to be prevalent at temperature around $10^{7}$ GeV.}
    \label{fig:Fig5}
\end{figure}
Let $T_{\mathrm{rad}}$ be attained at the cost of $N(T_{\mathrm{rad}})$ number of e-folds, then from Eqs. (\ref{eq:T_rad_2}) and (\ref{eq:e_folds}) 
\begin{equation}
    N(T_{\mathrm{rad}})=\ln{\left(\frac{T_{\mathrm{end}}}{T_{\mathrm{rad}}}\right)}=\frac{1}{2}\ln{\left(\frac{12\pi^3}{g^2}\right)}=12.1698.
\end{equation}
Therefore the radiation begins to dominate exactly at $12.1698$ e-folds after inflation at the temperature around $10^7$ GeV (see figure \ref{fig:Fig5}). Interestingly, this result matches with the one at the `crossing point' of energy densities in figure \ref{fig:Fig2}.\par Up to now, it is easily realizable that, SB would occur at any temperature between $\sim 10^{11}-10^{7}$ GeV with the number of e-folds lying in the range $1.47-12.17$. However, the exact value of this temperature can be estimated from the freeze-out temperature $T_F$, which is linked with the BTER, calculated in next sub-section.\par To get a rough estimate of $T_F$, if we assume that at the freeze-out point $\eta_F$ is equal to the experimental value $\eta=8.6\times 10^{-11}$ in Eq. (\ref{eq:BTER_final}), $\frac{\lambda'M_p}{M}$ is restricted between $1$ and $8$ according to Carroll bound, say, $\frac{\lambda'M_p}{M}=5$ and $T_{\mathrm{th}}=8.5\times 10^{11}$
 GeV, taken from figure \ref{fig:Fig4}, then within the model concerned $T_F$ can be calculated to be $1.69\times 10^8$ GeV, which is above $T_{\mathrm{rad}}$ ($\sim 10^7$ GeV, see figure \ref{fig:Fig5}). Now, in the next sub-section it will be interesting to look at the conditions in which $T_F$ acquires the value of this order.
 \subsection{Calculations of \texorpdfstring{\boldmath$T_F$}{} and \texorpdfstring{$\eta_F$}{}}
In order to compute the freeze-out temperature $T_F$, we consider a $B$-violating and $(B-L)$-conserving process, of which the rate, derived from an SM four-fermion non-renormalizable point interaction \cite{DeFelice:2002ir,Trodden:2003yn,Ahmad:2019jbm,Basak:2021cgk,Kolb:1990vq}, can be written as,   
\begin{equation}
    \Gamma_{B-L} (T)=\frac{\Bar{g}^2}{\Tilde{M}^4}T^5,
    \label{eq:rate}
\end{equation}
where $\Bar{g}<1$ is the coupling constant and $\Tilde{M}$ indicates the effective cut-off of the said interaction. At the specific temperature, $T_F$, the rate $\Gamma_{B-L}$ becomes of the same order as that of the Hubble expansion rate $H_{\mathrm{kin}}$ of the universe \emph{i.e.} $\Gamma_{B-L} (T_F)\sim H_{\mathrm{kin}}(T_F)$ (called the \emph{freeze-out condition}) and beyond that no $B$-violation takes place. Consequently, the baryon number saturates at the freeze-out value. Now, using Eqs. (\ref{eq:H_kin}), (\ref{eq:T_end_2}) and (\ref{eq:rate}), we obtain
\begin{equation}
    T_F=\frac{H_{\mathrm{end}}^{1/2}\Tilde{M}^2}{T_{\mathrm{end}}^{3/2}\Bar{g}}.
    \label{eq:tfreeze}
\end{equation}
\begin{figure}[H]
    \begin{subfigure}{0.5\linewidth}
  \centering
\includegraphics[width=70mm,height=55mm]{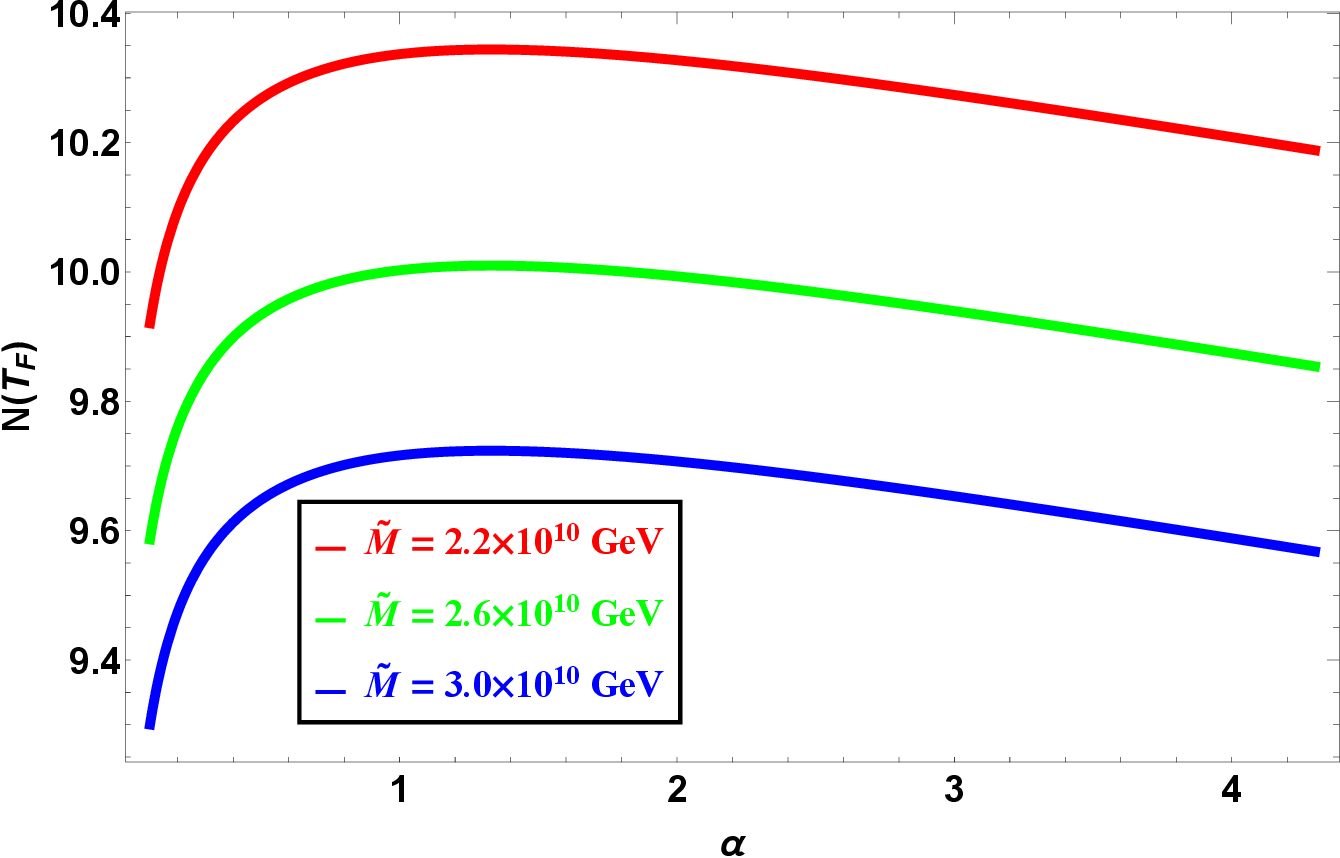}
  \subcaption{}
   \label{fig:Fig6a}
\end{subfigure}
\begin{subfigure}{0.5\linewidth}
  \centering
\includegraphics[width=70mm,height=55mm]{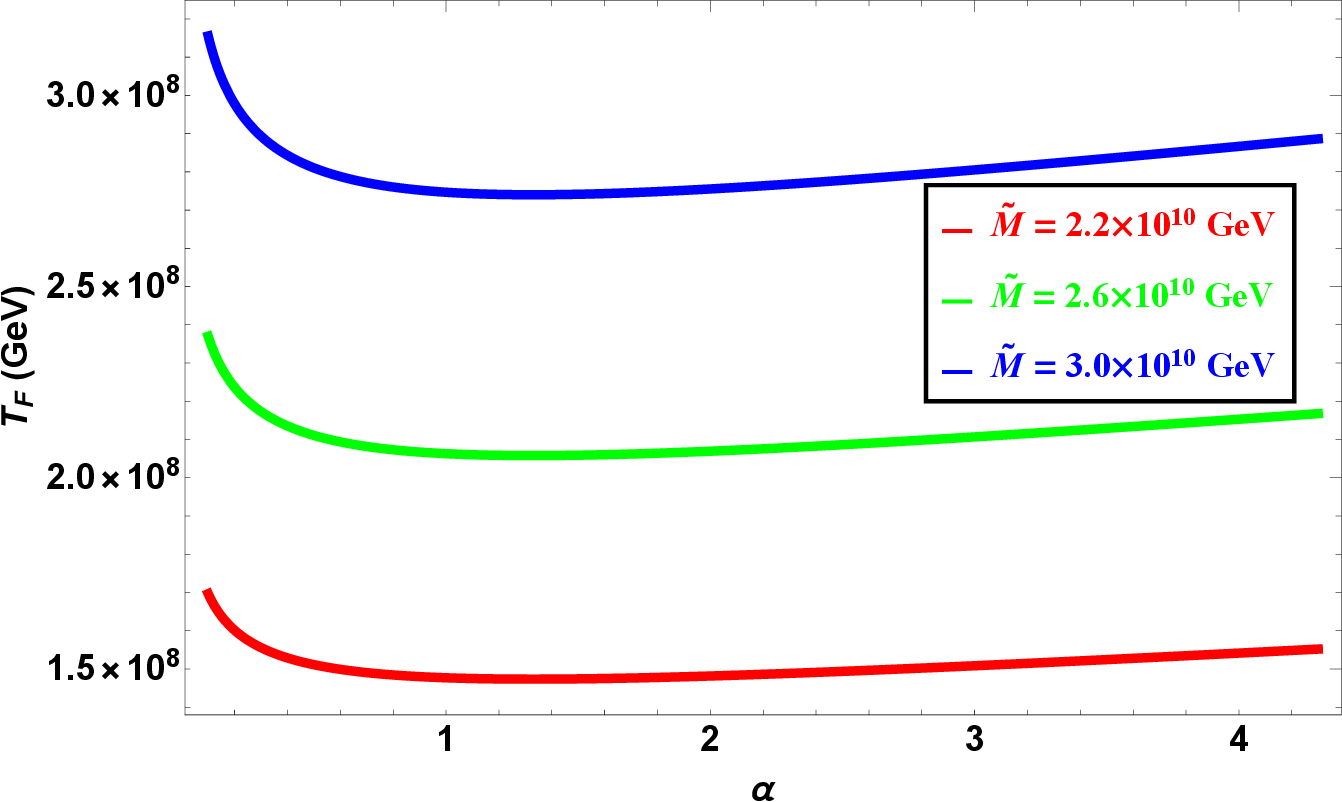}
   \subcaption{}
    \label{fig:Fig6b}
\end{subfigure}%
\vspace{0.05\linewidth}
\begin{subfigure}{0.5\linewidth}
  \centering
\includegraphics[width=70mm,height=55mm]{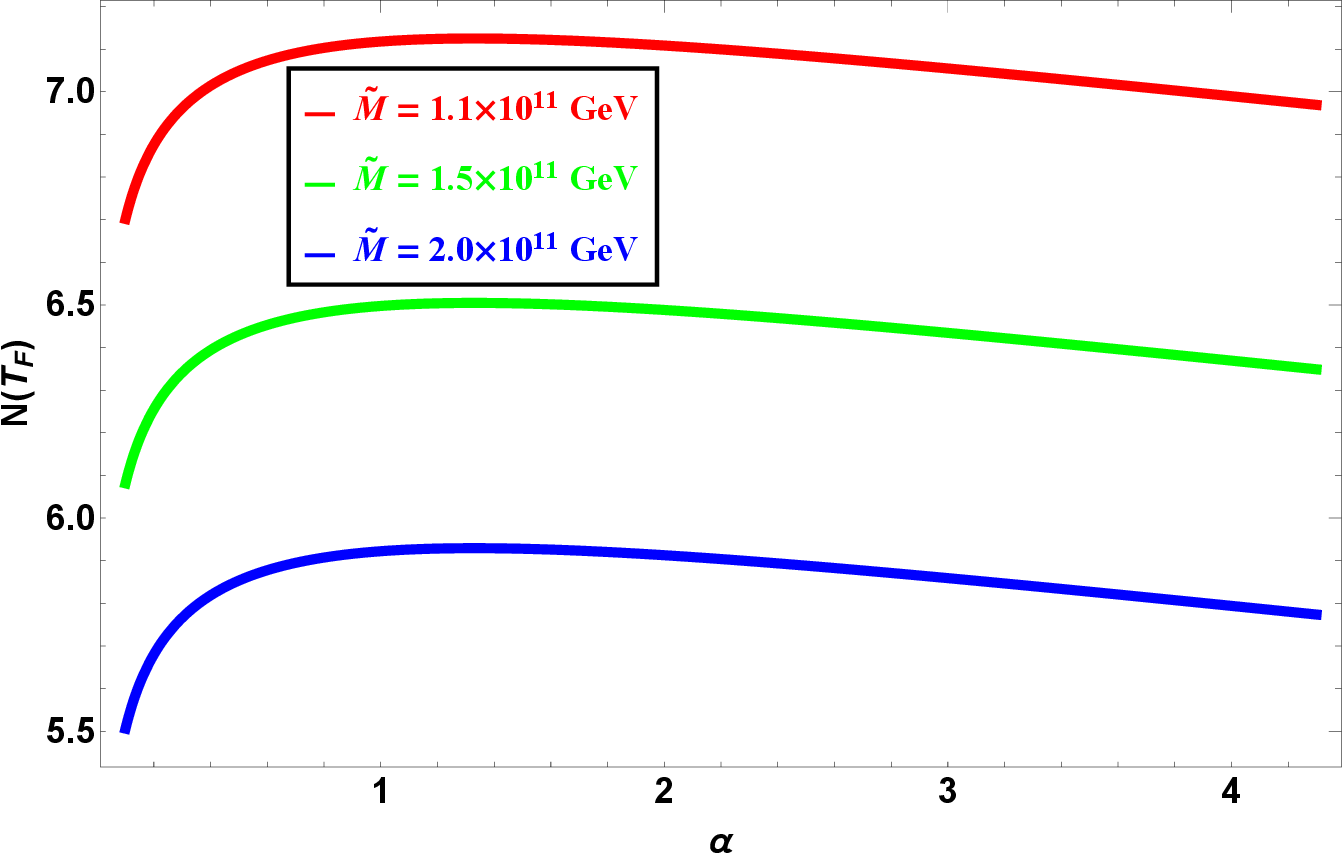}
   \subcaption{}
    \label{fig:Fig6c}
\end{subfigure}
\begin{subfigure}{0.5\linewidth}
  \centering
\includegraphics[width=70mm,height=55mm]{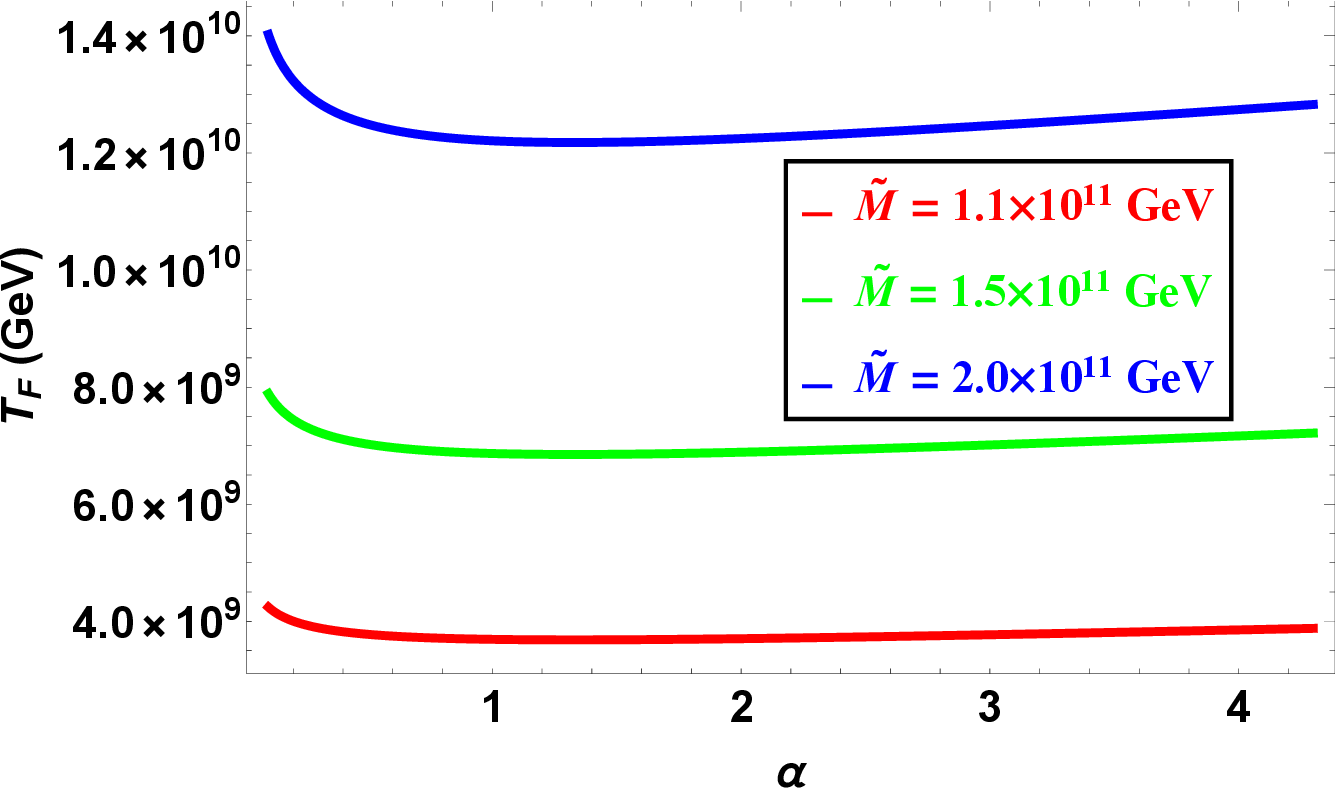}
   \subcaption{}
    \label{fig:Fig6d}
\end{subfigure}
     \caption{\emph{Left column}: Number of e-folds elapsed to attain the freezing of baryon non-conserving process at different mass scales. \emph{Right column}: The variations of freeze-out temperatures with $\alpha$ show that, freezing happens at $\sim 10^8-10^{10}$ GeV for a given value of $\Tilde{M}$ and $\Bar{g}=0.8$. For a particular value of $\alpha$, larger value of number of e-folds corresponds to smaller value of $T_F$, which is consistent with the evolution of temperature of the universe with its expansion, described in previous sub-section. With the increase of the mass scale $\Tilde{M}$, the number of e-folds decreases and $T_F$ increases for a given value of $\alpha$, which are also clear from Eqs. (\ref{eq:tfreeze}) and (\ref{eq:tf_efolds}).}
    \label{fig:Fig6}
\end{figure}
The required number of e-folds to reach the temperature $T_F$ can be calculated from Eqs. (\ref{eq:T_end_2}) and (\ref{eq:e_folds}) (as done before in the cases of $N(T_{\mathrm{
th}})$ and $N(T_{\mathrm{rad}})$) as, 
\begin{equation}
    N(T_F)=\ln{\left(\frac{T_{\mathrm{end}}}{T_F}\right)}. 
    \label{eq:tf_efolds}
\end{equation}
In figure \ref{fig:Fig6} the freeze-out temperature $T_F$ (see Eq. (\ref{eq:tfreeze})) and the corresponding number of e-folds $N(T_F)$ are plotted as functions of $\alpha$ for various values of the mass-scale $\Tilde{M}$. Here, $\Tilde{M}$ is so chosen that $T_F$ could lie between $T_{\mathrm{th}}$ ($\sim 10^{11}$ GeV) and $T_{\mathrm{rad}}$ ($\sim 10^7$ GeV), as stated in previous sub-section. Figure \ref{fig:Fig6} highlights that, depending upon the values of $\Tilde{M}$, the freezing of $B$-violating process occurs at a temperature around $10^8-10^{10}$ GeV when the universe has covered $5.5-10.3$ number of e-folds, which lie in the midst of thermalization and radiation domination.\par 
\begin{figure}[H]
    \begin{subfigure}{0.5\linewidth}
  \centering
\includegraphics[width=70mm,height=55mm]{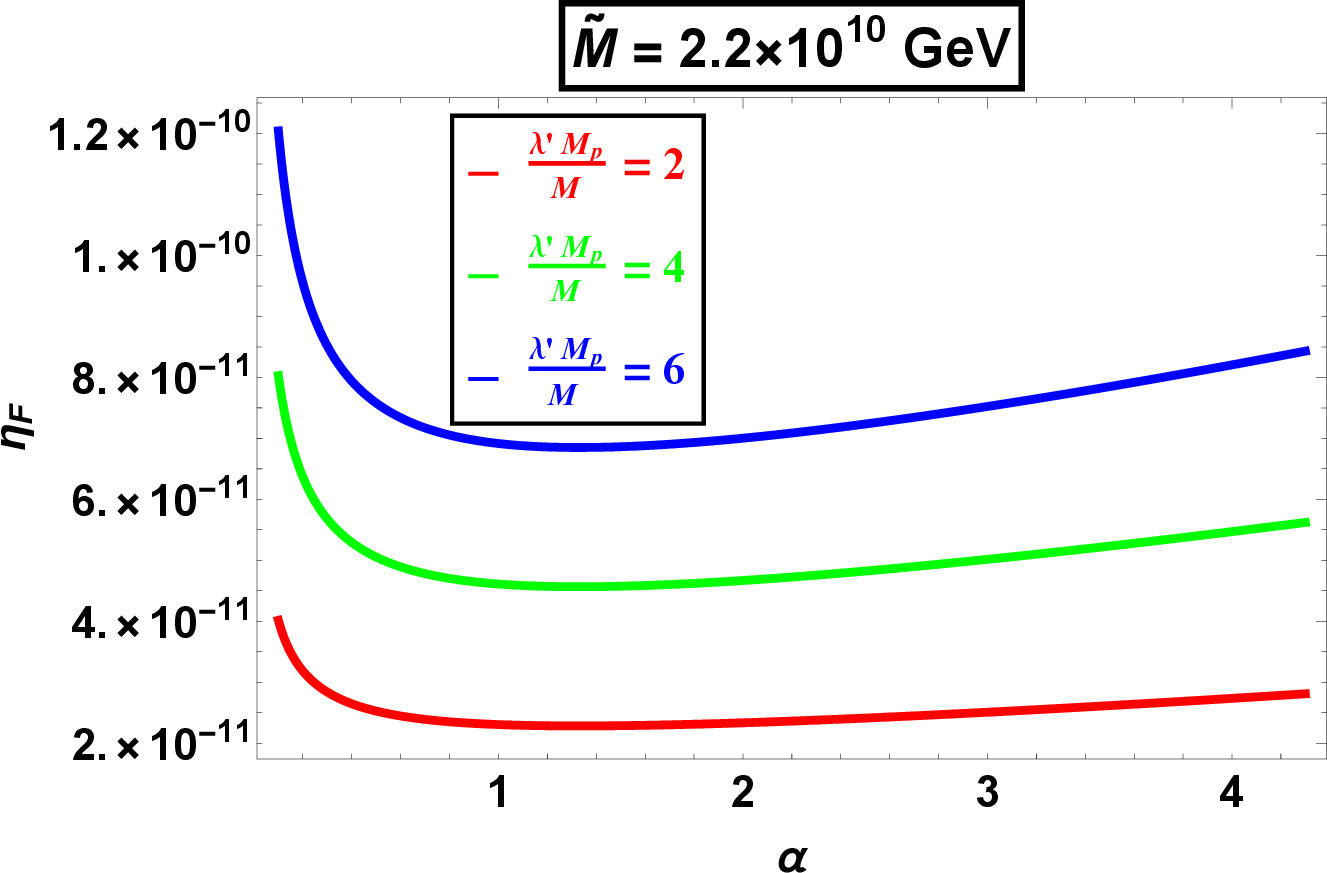}
  \subcaption{}
   \label{fig:Fig7a}
\end{subfigure}
\begin{subfigure}{0.5\linewidth}
  \centering
\includegraphics[width=70mm,height=55mm]{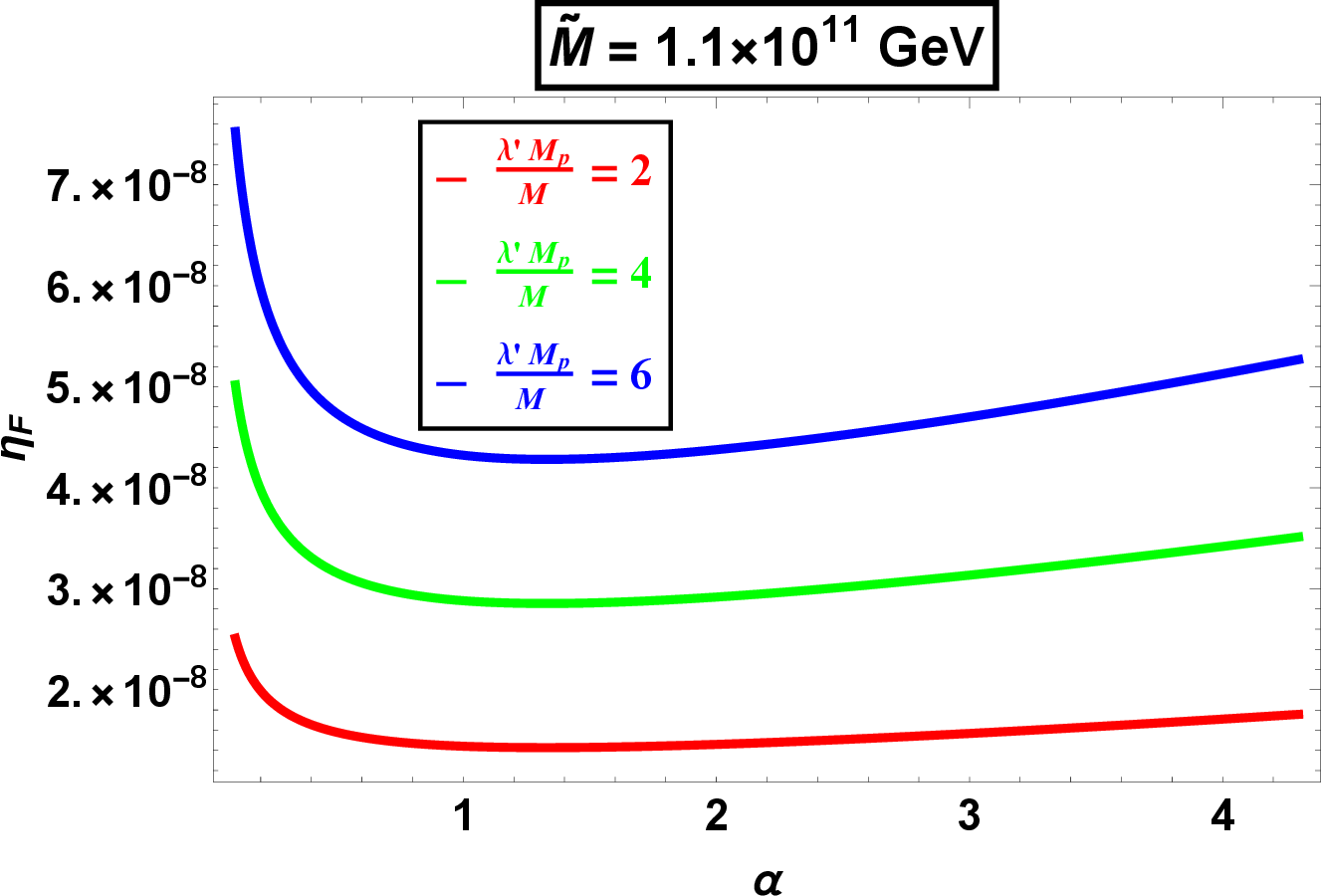}
   \subcaption{}
    \label{fig:Fig7b}
\end{subfigure}%
\vspace{0.05\linewidth}
\begin{subfigure}{0.5\linewidth}
  \centering
\includegraphics[width=70mm,height=55mm]{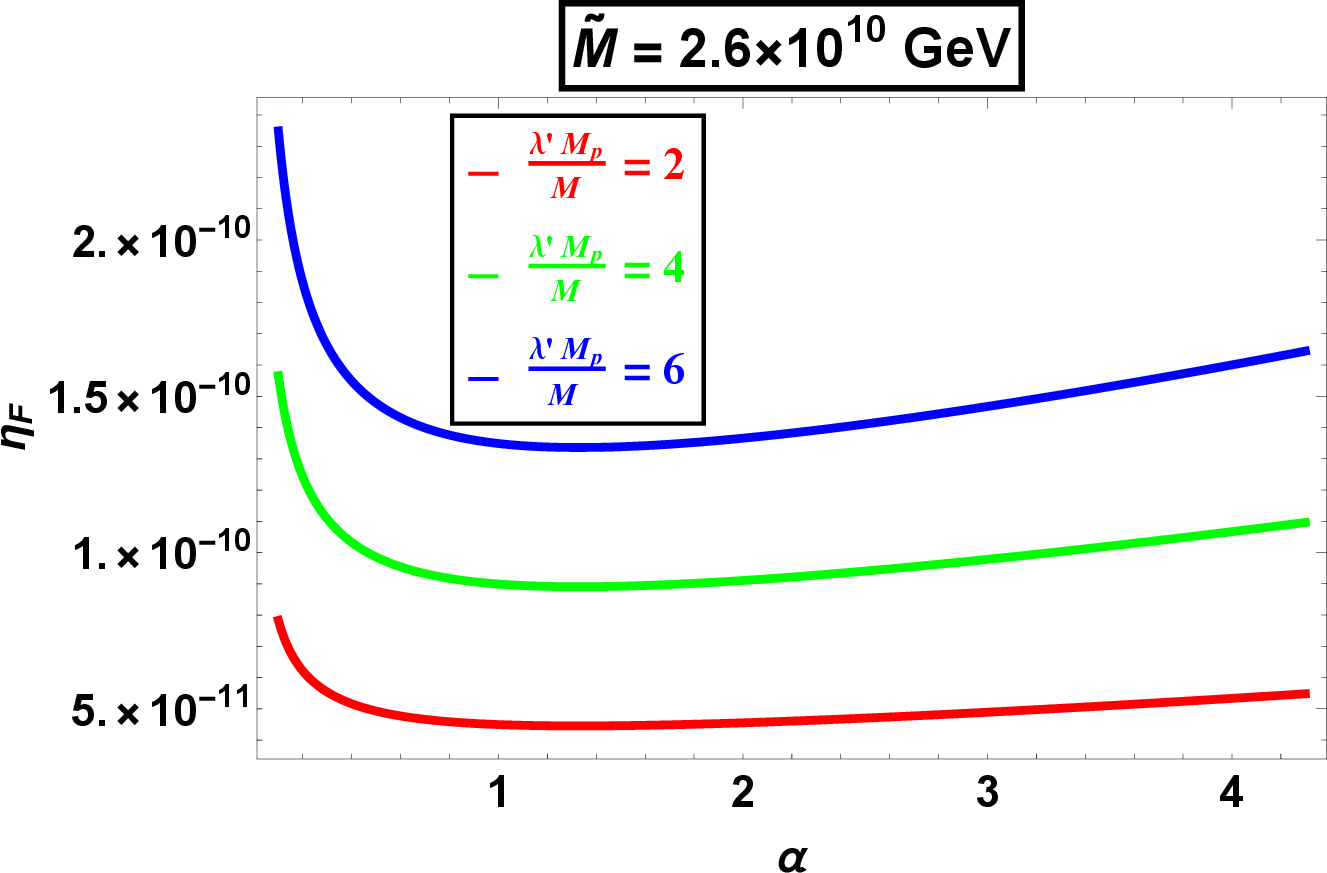}
   \subcaption{}
    \label{fig:Fig7c}
\end{subfigure}
\begin{subfigure}{0.5\linewidth}
  \centering
\includegraphics[width=70mm,height=55mm]{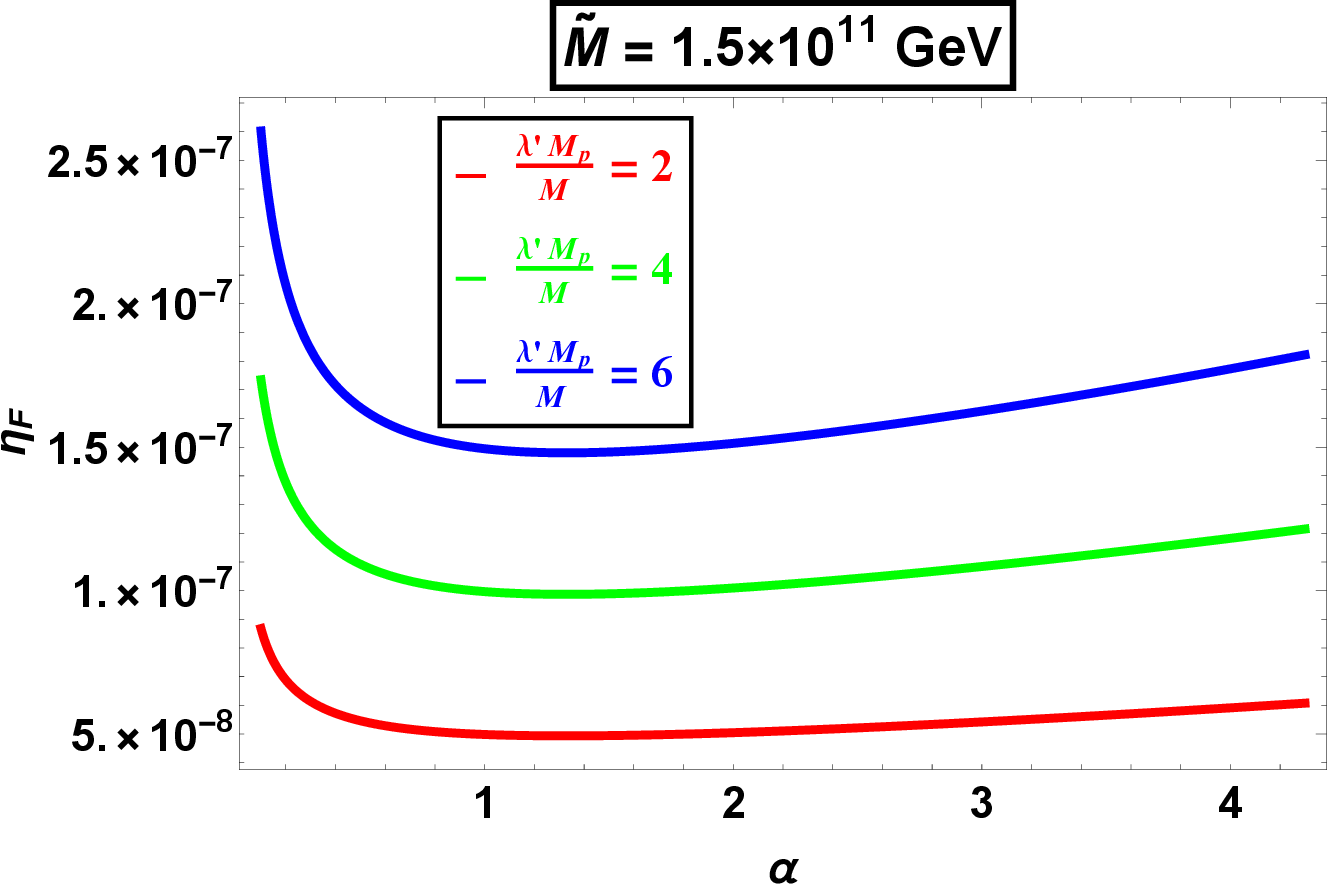}
   \subcaption{}
    \label{fig:Fig7d}
\end{subfigure}%
\vspace{0.05\linewidth}
\begin{subfigure}{0.5\linewidth}
  \centering
\includegraphics[width=70mm,height=55mm]{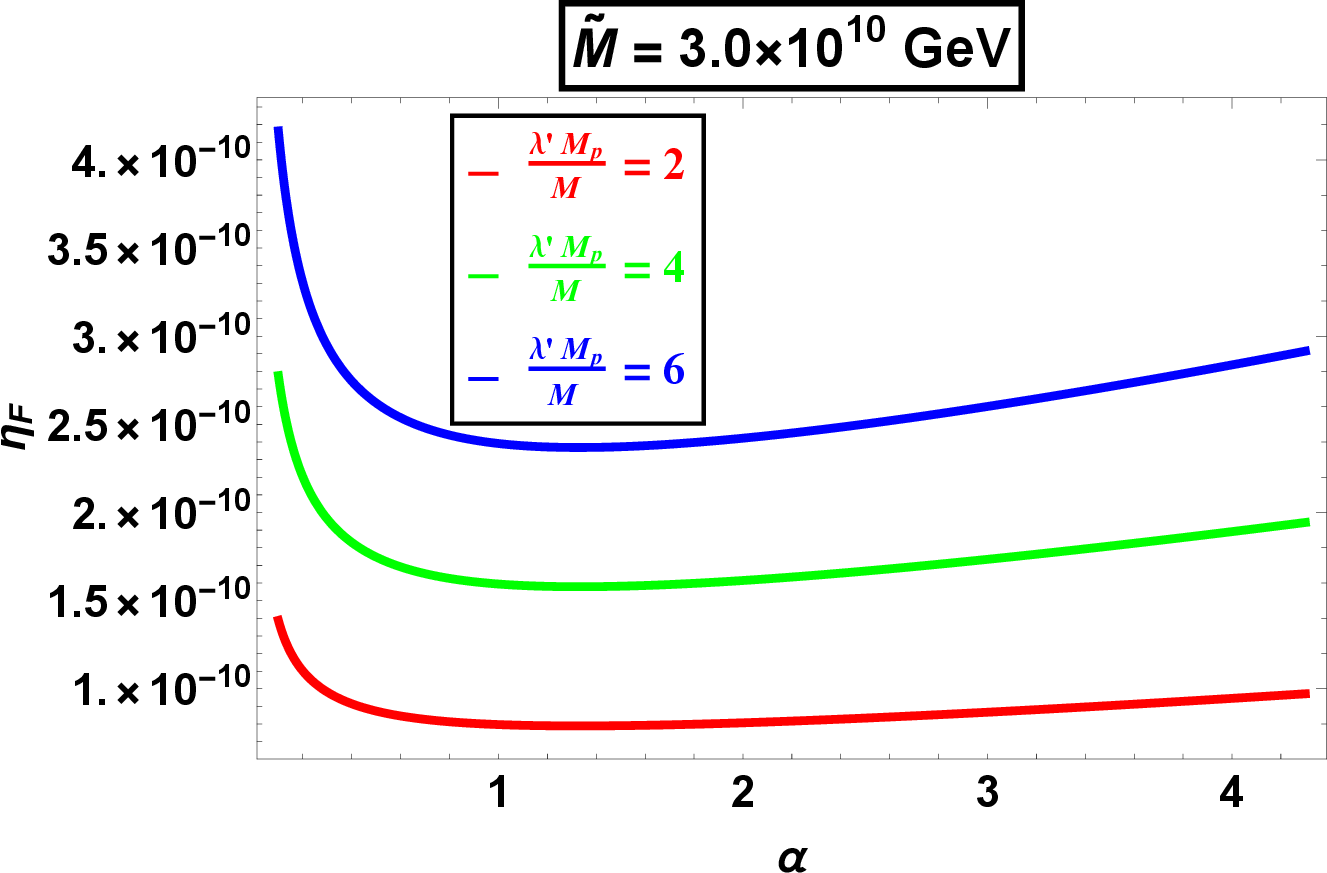}
   \subcaption{}
    \label{fig:Fig7e}
\end{subfigure}
\begin{subfigure}{0.5\linewidth}
  \centering
\includegraphics[width=70mm,height=55mm]{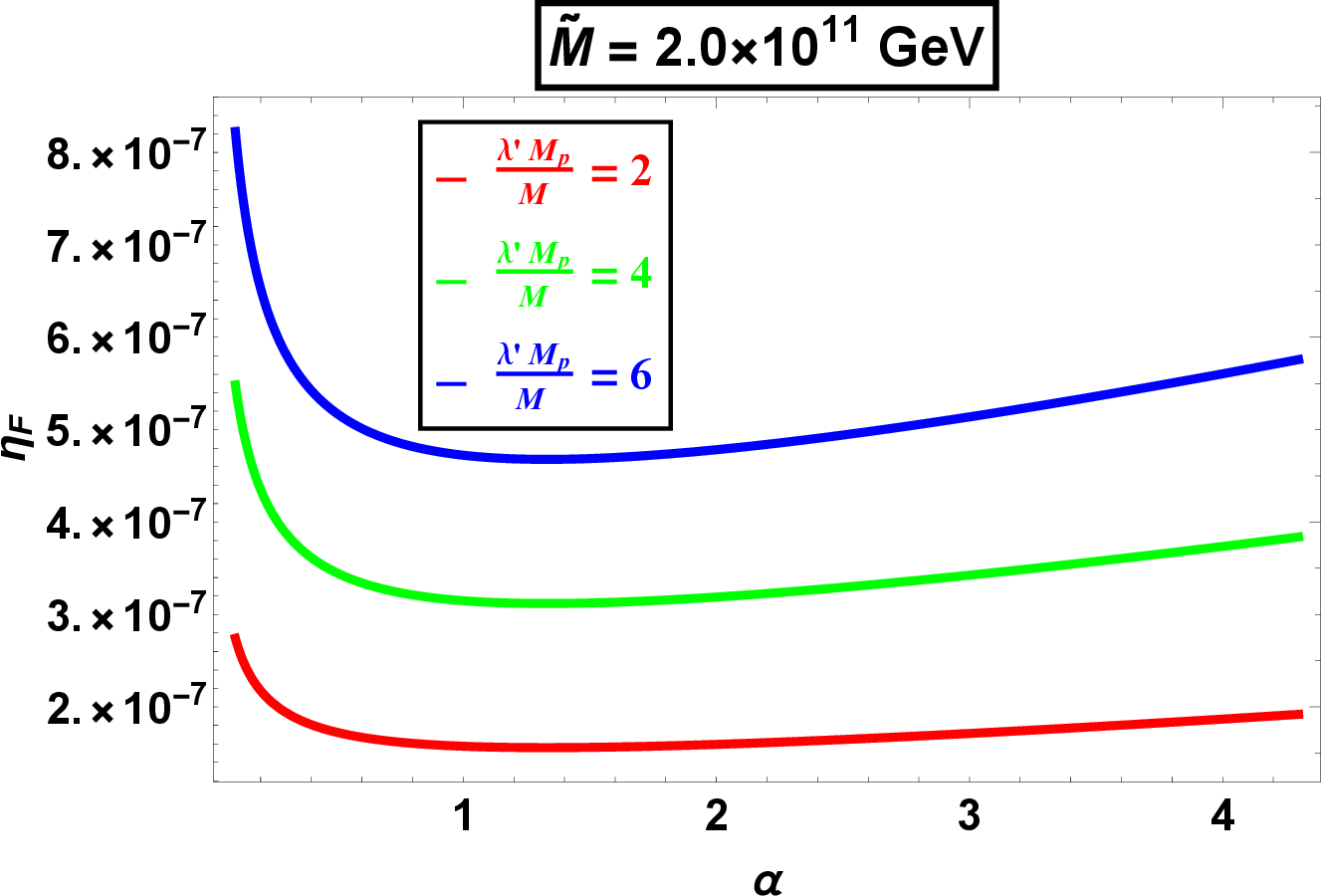}
   \subcaption{}
    \label{fig:Fig7f}
\end{subfigure}
     \caption{Variations of freeze-out values of BTER with $\alpha$ for three values of $M$ and two sets of values of $\Tilde{M}$ corresponding to $\sim10^{10}$ GeV and $\sim10^{11}$ GeV. For particular values of $\alpha$ and $\Tilde{M}$, $\eta_F$ decreases with the increase in $M$. Similarly, for particular values of $\alpha$ and $M$, $\eta_F$ increases with the increase in $\Tilde{M}$. These also become clear from Eq. (\ref{eq:Mod_eta_F}).}
    \label{fig:Fig7}
\end{figure}
Finally, the task is to estimate the freeze-out value of BTER, $\eta_F$, from Eq. (\ref{eq:BTER_final}). We can even simplify that by putting the expressions of $T_F$ from Eq. (\ref{eq:tfreeze}) and $T_{\mathrm{th}}$ from Eq. (\ref{eq:thermal_T}) in Eq. (\ref{eq:BTER_final}) as
\begin{equation}
    \eta_F = 7.25\times 10^{15}\times \left(\frac{\lambda' M_p}{M}\right)\times\frac{\Tilde{M}^4}{M_p^2\sqrt{V_{\mathrm{end}}}}=7.25\times 10^{15} \left(\frac{\lambda' M_p\Tilde{M}^4}{M M_p^2}\right)\times\frac{e^{\frac{n}{2}\left[1-\sqrt{1-\left(\frac{\sqrt{12\alpha}}{n}\right)}\right]}}{\sqrt{V_{*}(n,\alpha)}},
    \label{eq:Mod_eta_F}
\end{equation} where $V_{\mathrm{end}}$ satisfies Eq. (\ref{eq:Vend}). To compute $\eta_F$ as function of $\alpha$, three values of $M$ \emph{viz.,} $M=\lambda'M_p/2=1.2\times 10^{18}$ GeV, $\lambda'M_p/4=6.07\times 10^{17}$ GeV, $\lambda'M_p/6=4.05\times 10^{17}$ GeV for $\lambda'\approx 1$ and the estimated values of $T_{\mathrm{th}}$ and $T_F$ shown in figures \ref{fig:Fig4} and \ref{fig:Fig6}, respectively, have been considered. The choices of $\Tilde{M}$ remain the same as shown in figure \ref{fig:Fig6}. Two sets of $\Tilde{M}$, each containing  three values, corresponding to $\Tilde{M}\sim 10^{10}$ GeV and $\Tilde{M}\sim 10^{11}$ GeV are taken into account.\par In figure \ref{fig:Fig7}, the obtained values of $\eta_F$ with the variation of $\alpha$ from $\alpha=0.1$ to $\alpha=4.3$ are shown for $n=122$. The left column corresponds to the variations when $\Tilde{M}\sim 10^{10}$ GeV and the right column comprises that for $\Tilde{M}\sim 10^{11}$ GeV. The results show that the cut-off scale $M$ has no such significant effects, where as, $\Tilde{M}$ affects $\eta_F$ significantly. When $\Tilde{M}\sim 10^{10}$ GeV, $\eta_F$ is of the order of $10^{-10}-10^{-11}$, which is required for obtaining the experimental bound. Similarly, when $\Tilde{M}\sim 10^{11}$ GeV, $\eta_F$ is of the order of $10^{-7}-10^{-8}$, which is higher than that of the experimental bound.\par At $\Tilde{M}=2.2\times 10^{10}$ GeV and $M=4.05\times 10^{17}$ GeV (see the blue curve of figure \ref{fig:Fig7a}), the $\eta_F$ is found to be closest to the experimental value, $\eta\approx (8.6\pm 0.1)\times 10^{-11}$. The values of $\eta_F$ have been found within the range $\eta_F=8.7\times 10^{-11} - 8.5\times 10^{-11}$ for $\alpha=0.28 - 0.30$ respectively. Therefore, in the lower end of $\alpha$ \textit{viz.,} $\alpha=0.28 - 0.30$, the calculated BTER satisfies the experimental value with specified error bar for the mass scales $\Tilde{M}=2.2\times 10^{10}$ GeV and $M=4.05\times 10^{17}$ GeV.\par
\begin{table}[ht]
\caption{Calculated values of $\eta_F$, $T_{\mathrm{th}}$, $T_F$, $T_{\mathrm{rad}}$ and $T_{\mathrm{end}}$ for the given range of values of $\alpha$, where $\eta_F$ satisfies the experimental requirement: $\eta\approx (8.6\pm 0.1)\times 10^{-11}$.}
    \centering
    \begin{tabular}{|c|c|c|c|c|c|c|}
    \hline
       $\alpha$ & $\eta_F/(10^{-11})$ & $T_{\mathrm{th}}/(10^{11})$ & $T_F/(10^{8})$ & $\frac{T_{\mathrm{th}}}{T_F}\times 10^{-3}$ & $T_{\mathrm{rad}}/(10^{7})$ & $T_{\mathrm{end}}/(10^{12})$ \\
       \hline
       \hline
        $0.28 - 0.30$ & $8.7 - 8.5$ & $8.63 - 8.64$ & $1.56 - 1.55$ & $5.532 - 5.574$ & $2.11 - 2.13$ & $4.05 - 4.10$\\
        \hline
    \end{tabular}
    \label{tab:tab_1}
\end{table}
Table \ref{tab:tab_1} shows that, in the regime, where $\eta_F$ satisfies the experimental value, the maximum thermalization temperature, $T_{\mathrm{th}}$ and the freeze-out temperature, $T_F$ are found to be $T_{\mathrm{th}}= 8.63\times 10^{11} - 8.64\times 10^{11}$ GeV and $T_F=1.56\times 10^{8} - 1.55\times 10^{8}$ GeV for $\alpha=0.28 - 0.30$ (see figure \ref{fig:Fig4} and the red curve of figure \ref{fig:Fig6b}). Here, $T_{\mathrm{th}}/T_F=5.532\times 10^3 - 5.574 \times 10^3$, which are close to the value given in Ref. \cite{Ahmad:2019jbm}. In this range of $\alpha$, the temperature at the end of inflation (or the start of kination) $T_{\mathrm{end}}$ and the temperature of radiation domination, $T_{\mathrm{rad}}$, are obtained as, $T_{\mathrm{end}} = 4.05\times 10^{12} - 4.10\times 10^{12}$ GeV and $T_{\mathrm{rad}}=2.11\times 10^7 - 2.13\times 10^7$ GeV (see figures \ref{fig:Fig3} and \ref{fig:Fig5}). Also, the symmetry breaking scale $M$ ($\sim 10^{17}$ GeV) is found to be $10^5$ times larger than the Hubble rate $H_{\mathrm{end}}$ ($\sim 10^{12}$ GeV) (see figure \ref{fig:Fig3}), which is an important result according to Ref. \cite{DeSimone:2016ofp}.\par Thus, from all the analyses done so far, it can be said that, the SB takes place at the temperature $T$ lying within the following inequality
\begin{equation}
\begin{split}
    &T_{\mathrm{rad}} (\sim 10^7 \mathrm{GeV})<T_F (\sim 10^8 \mathrm{GeV})<T (\mathrm{GeV})<T_{\mathrm{th}} (\sim 10^{11} \mathrm{GeV})<T_{\mathrm{end}} (\sim 10^{12} \mathrm{GeV})\\&<M (\sim 10^{17} \mathrm{GeV})<M_p (\sim 10^{18} \mathrm{GeV}).
\end{split}
\end{equation}
\subsection{Spectra of the relic gravitational waves}
\label{sec:Gravity_waves}
\begin{figure}[H]
	\centering
\includegraphics[width=0.8\linewidth]{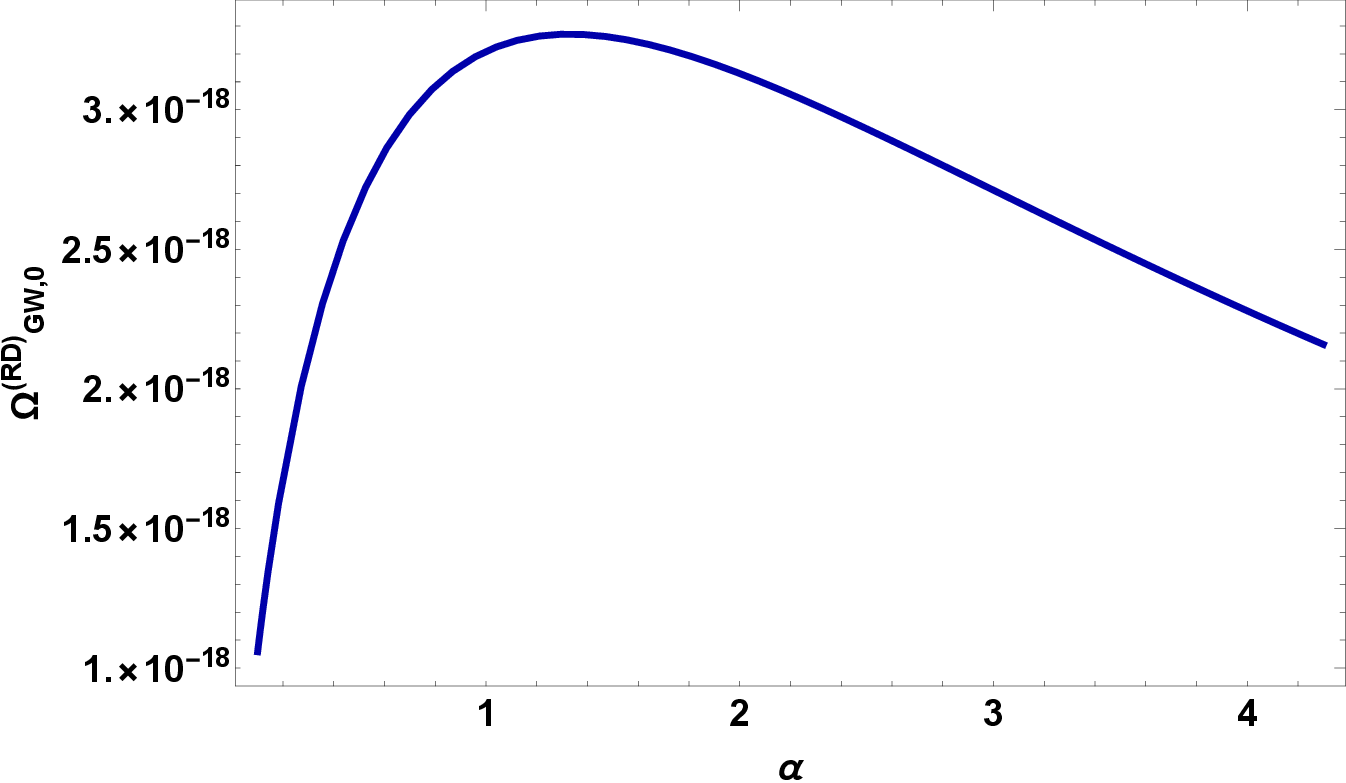}
	\caption{Plot of the RMS values of the amplitudes for the present-day RD-GW spectrum against $\alpha$.}
	\label{fig:Fig8}
 \end{figure}

\begin{figure}[H]
	\centering
\includegraphics[width=0.8\linewidth]{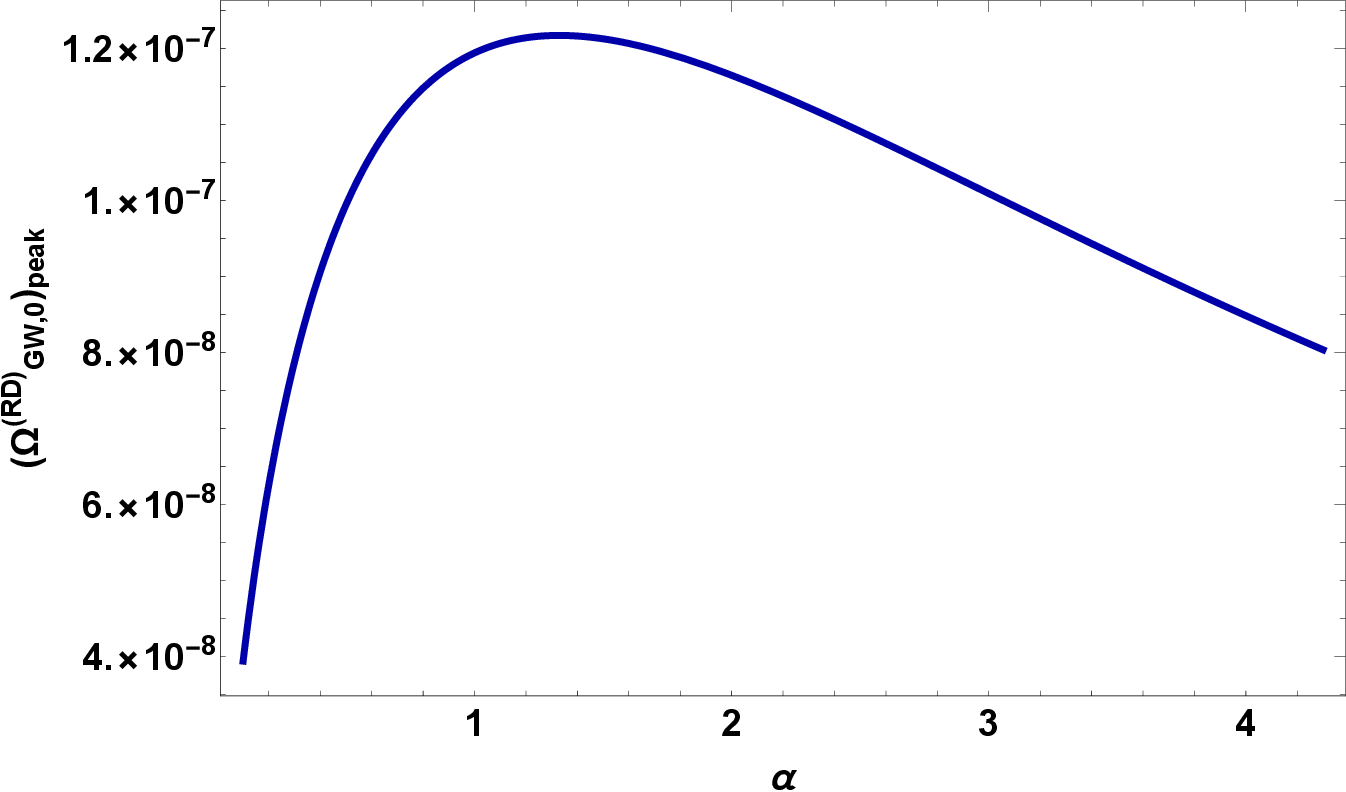}
	\caption{Plot of the peak values of the amplitudes for the present-day RD-GW spectrum against $\alpha$.}
	\label{fig:Fig9}
 \end{figure}

Gravitational waves (GWs) are the results of tensor perturbations in spatially-flat FLRW metric defined by (written in standard notations) \cite{Baumann:2009ds}
\begin{equation}
    ds^2=-dt^2+a^2(t)\left(\delta_{mn}+h_{mn}\right)dx^m dx^n.
\end{equation} $h_{mn}$ are the metric perturbations, which are transverse ($\partial_m h_{mn}=0$) having two crossed polarization states and trace-less ($h_{mm}=0$). The root-mean-square (RMS) value of the amplitudes of the GW spectrum, $\Omega_{GW} (k,\tau)$, is measured as the ratio of logarithmic derivative of GW energy density $\rho_{GW}$ with respect to the mode $k$ and the conformal time ($\tau$)-dependent critical energy density $\rho_c=3M_p^2 H^2(\tau)$ \cite{Ahmad:2017itq,Basak:2021cgk,Das:2023nmm}.
This can be transferred to its present-day value $\Omega_{GW,0}$ \cite{Ahmad:2017itq,Basak:2021cgk,Das:2023nmm} by extracting the effects of (dimensionless) inflationary tensor power spectrum $\Delta^2_h(k)$ (see \cite{Sarkar:2021ird} for the calculations of mode-dependent tensor power spectrum) at present-day, through a transfer function $T_{\mathrm{trans}}^2(k,\tau)$ \cite{Watanabe:2006qe,Kuroyanagi:2008ye,Boyle:2005se}. This function encodes the post-inflationary evolution of a mode $k$ from horizon re-entry to time ($\tau$) of observation. Now, a number of expressions of $\Omega_{GW,0}$ exist depending upon the various epochs in which the modes re-enter the horizon. Among them, only the kinetic-dominated (KD) and radiation-dominated (RD) epochs are the subjects of interest in the present work. They are given by \cite{WaliHossain:2014usl,Ahmad:2017itq,Ahmad:2019jbm,Basak:2021cgk,Das:2023nmm},
\begin{equation}
    \Omega^{(\mathrm{RD})}_{GW,0} (k)=\left(\frac{\Omega_{\mathrm{rad},0}}{6\pi^2}\right)\left(\frac{H_{\mathrm{inf}}^2}{M_p^2}\right)\left(\frac{g_{*}}{g_{*,0}}\right)\left(\frac{g_{*s,0}}{g_{*s}}\right)^{4/3}
    \label{eq:RD_strain}
\end{equation}
for $k_{\mathrm{equal}}<k\leq k_{\mathrm{rad}}$
and 
\begin{equation}
    \Omega^{(\mathrm{KD})}_{GW,0} (k)=\Omega^{(\mathrm{RD})}_{GW,0}(k)\left(\frac{k}{k_{\mathrm{rad}}}\right)
    \label{eq:KD_strain}
\end{equation}
for $k_{\mathrm{rad}}<k\leq k_{\mathrm{end}}$. $\Omega_{\mathrm{rad},0}=9.13\times 10^{-5}$ \cite{Planck:2018jri,Planck:2018vyg} is the present-day density parameter of radiation. $g_{*s}$ and $g_{*}$ are relativistic degrees of freedom corresponding to entropy density and radiation density respectively, which are of the same order \emph{i.e.} $10^{2}$ \cite{Basak:2021cgk}. $g_{*s,0}=3.91$ and $g_{*,0}=3.36$ are their present values \cite{Planck:2018jri,Planck:2018vyg}. The subscripts `end', `rad' and `equal' in $k$ refer to the modes taking re-entry during the end of inflation (or start of kination), radiation era and matter-radiation equality, respectively. Here, $H_{\mathrm{inf}}$ refers to the inflationary Hubble parameter which is generically mode-dependent (see Ref. \cite{Sarkar:2021ird}) and hence a little complicated. But it can be simplified for the plateau-type potential considered here (see Eq. (\ref{eq:Qpot})), for which tensor-to-scalar ratio is small, by approximating $H_{\mathrm{inf}}\approx H_{\mathrm{end}}$, which is $k$-independent. Therefore the GW spectrum during the radiation era is roughly scale-independent. It depends only on $\alpha$ through $H_{\mathrm{end}}$, while that for kinetic domination is slightly blue-tilted, which will be examined in the next paragraph. In figure \ref{fig:Fig8}, we plot the RMS values of the amplitudes of the present-day RD-GW spectrum against $\alpha$, and the results show that they are $\sim 10^{-18}$, which is indeed very small, lying within the projected ranges of various ongoing experiments (see figure \ref{fig:Fig11}).
 The corresponding peak value \cite{Ahmad:2017itq,Basak:2021cgk,Ahmad:2019jbm} is given by,
 \begin{equation}
\left(\Omega^{RD}_{GW,0}\right)_{\mathrm{peak}}=\Omega^{RD}_{GW,0}\left(\frac{T_{\mathrm{end}}}{T_{\mathrm{rad}}}\right)^2.
 \end{equation}
 In figure \ref{fig:Fig9} the peak values of amplitudes of present-day RD-GW spectrum are plotted against $\alpha$, in which the peak values are found to be $\sim 10^{-8}$, which satisfy the following constraint of BBN \cite{Cyburt:2015mya,Figueroa:2018twl}
 \begin{equation}
\left(\Omega^{RD}_{GW,0}\right)_{\mathrm{peak}}< 2.46\times 10^{-6}.
 \end{equation}
 For $\alpha=0.28 - 0.30$ we get $\Omega^{(\mathrm{RD})}_{GW,0}=2.05\times 10^{-18} - 2.10\times 10^{-18}$ and $\left(\Omega^{RD}_{GW,0}\right)_{\mathrm{peak}}=7.65\times 10^{-8} - 7.99\times 10^{-8}$.\par
 Therefore, the results found so far are viable in making a reliable connection between the GeV-scale - (spontaneous) baryogenesis and the BBN.\par 
The GW frequencies \cite{WaliHossain:2014usl,Ahmad:2017itq} having wave vectors $k_{\mathrm{rad}}$ and $k_{\mathrm{end}}$ indicated in Eqs. (\ref{eq:RD_strain}) and (\ref{eq:KD_strain}) can be calculated using, 
\begin{equation}
f_{\mathrm{rad}}=1600\times\left(\frac{T_{\mathrm{rad}}}{10^{10}\mathrm{GeV}}\right)\mathrm{Hz}
\end{equation} and 
\begin{equation}
    f_{\mathrm{end}}=8900\times\left(\frac{H_{\mathrm{inf}}}{g\times 10^{4}\mathrm{GeV}}\right)^{1/2}\mathrm{Hz}.
\end{equation}
\begin{figure}[H]
    \begin{subfigure}{0.5\linewidth}
  \centering
\includegraphics[width=70mm,height=55mm]{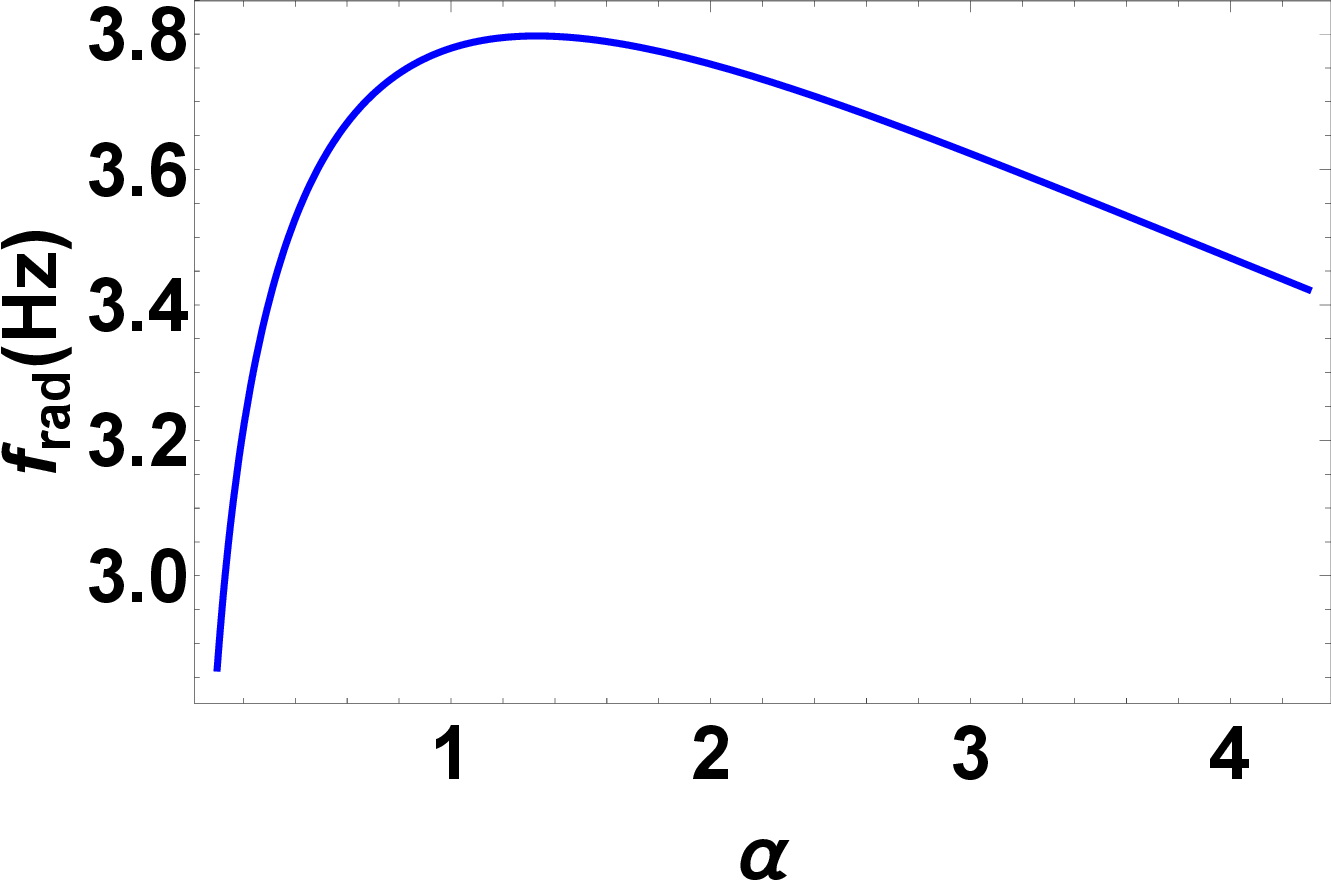}
  \subcaption{}
   \label{fig:Fig10a}
\end{subfigure}
\begin{subfigure}{0.5\linewidth}
  \centering
\includegraphics[width=70mm,height=55mm]{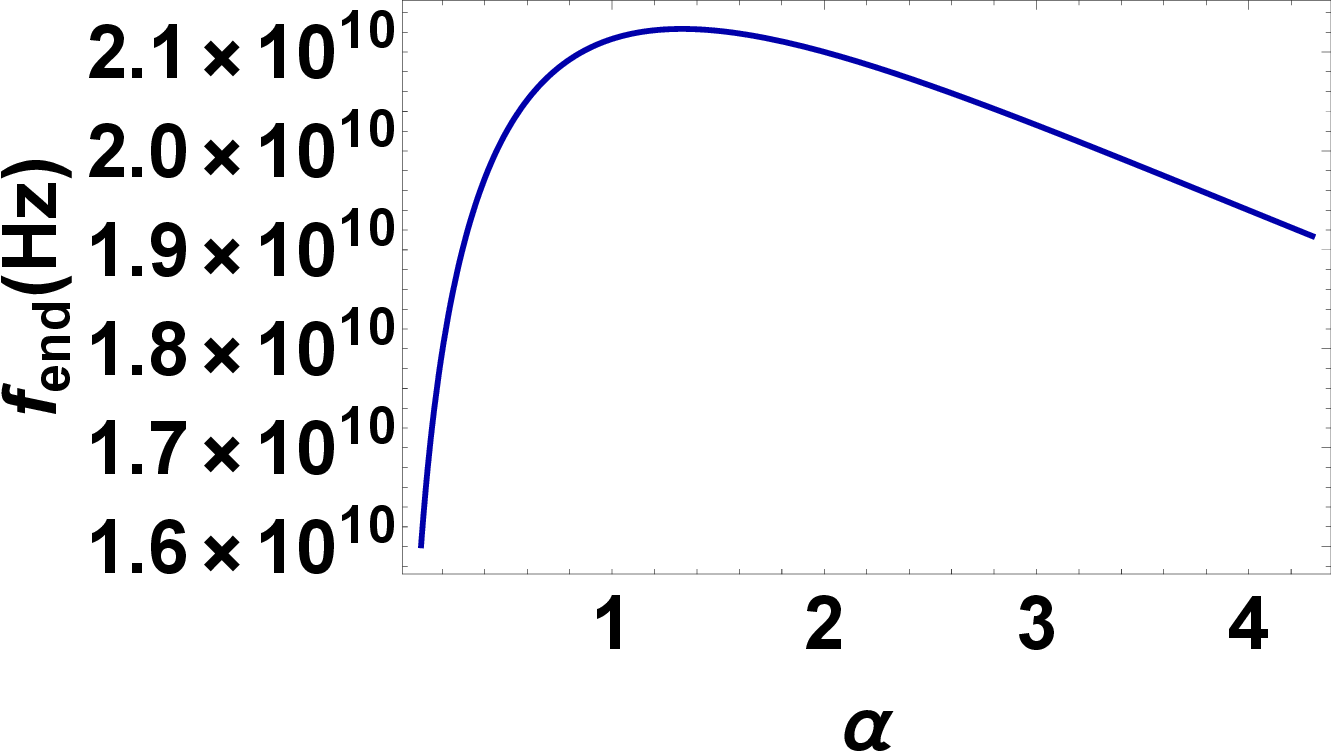}
   \subcaption{}
    \label{fig:Fig10b}
\end{subfigure}
     \caption{GW frequencies of the modes taking re-entry during radiation domination and the end of inflation (\emph{i.e.} start of kination).}
    \label{fig:Fig10}
\end{figure}
Figure \ref{fig:Fig10} shows that the obtained values of $f_{\mathrm{end}}$ and $f_{\mathrm{rad}}$ are $\sim 10^{10}$ Hz and $\sim 10^1$ Hz respectively. For $\alpha = 0.28 - 0.30$ we get $f_{\mathrm{end}} = 1.88\times 10^{10} - 1.91\times 10^{10}$ Hz and $f_{\mathrm{rad}} = 3.38 - 3.39$ Hz. The $f_{\mathrm{end}}$ frequencies are large enough and hence they are often called the \emph{blue tilted} \cite{Giovannini:1999qj,Ahmad:2017itq,Ahmad:2019jbm}. Ref. \cite{Ahmad:2019jbm} shows that the GW frequencies corresponding to the modes entering the horizon at present and during matter-radiation equality (corresponding to $k_{\mathrm{equal}}$) are of the order of $10^{-19}$ Hz and $10^{-17}$ Hz respectively, for which they are termed as the \emph{red tilted}.\par Therefore, during the transition from inflation to kination GWs of very high frequencies (blue tilted) are emitted. This is because a sharp transition  takes place from periods of inflation to kination due to a change of metric structure. Other transitions are smooth for which the corresponding emitted GWs are of low frequencies (red tilted). However, all GWs have very low amplitudes and therefore they are too feeble to detect.\par In the present era of precision cosmology and multi-messenger astronomy \cite{Bailes:2021tot}, several ground-based and space-based GW detectors are in operation to detect the GW frequencies from very low end to very high end (see figure \ref{fig:Fig11}). For example, Advanced Laser Interferometer Gravitational-Wave Observatory (Advanced-LIGO) \cite{LIGOScientific:2019vic}, Advanced-Virgo \cite{VIRGO:2014yos} and Kamioka Gravitational Wave Detector (KAGRA) \cite{KAGRA:2018plz} detect the GW-strain $\sim10^{-9}$ at roughly $100$ Hz. Again, the Laser Interferometer Space Antenna (LISA) \cite{LISA:2017pwj} and Deci-Hertz Interferometer Gravitational-wave Observatory (DECIGO) \cite{Kawamura:2011zz} are made to sense the frequencies approximately from $ 0.001$ Hz to $\sim 0.1$ Hz. Even weaker GWs of frequency like $\sim 10^{-8}$ Hz can be probed using Square Kilometer Array (SKA) \cite{Janssen:2014dka} (the largest radio telescope, so far) and recently, that of $\sim 10^{-9}$ Hz are being detected in North American Nanohertz Observatory for Gravitational Waves (NANOGrav) \cite{NANOGrav:2023gor,NANOGrav:2023hde,NANOGrav:2023hvm}, European Pulsar Timing Array (EPTA) + Indian Pulsar Timing Array (InPTA) \cite{EPTA:2023fyk,EPTA:2023gyr,EPTA:2023xxk}, Parkes Pulsar Timing Array (PPTA) \cite{Zic:2023gta,Reardon:2023gzh,Reardon:2023zen} and Chinese Pulsar Timing Array (CPTA) \cite{Xu:2023wog}.\par
 \begin{figure}[H]
	\centering
\includegraphics[width=0.8\linewidth]{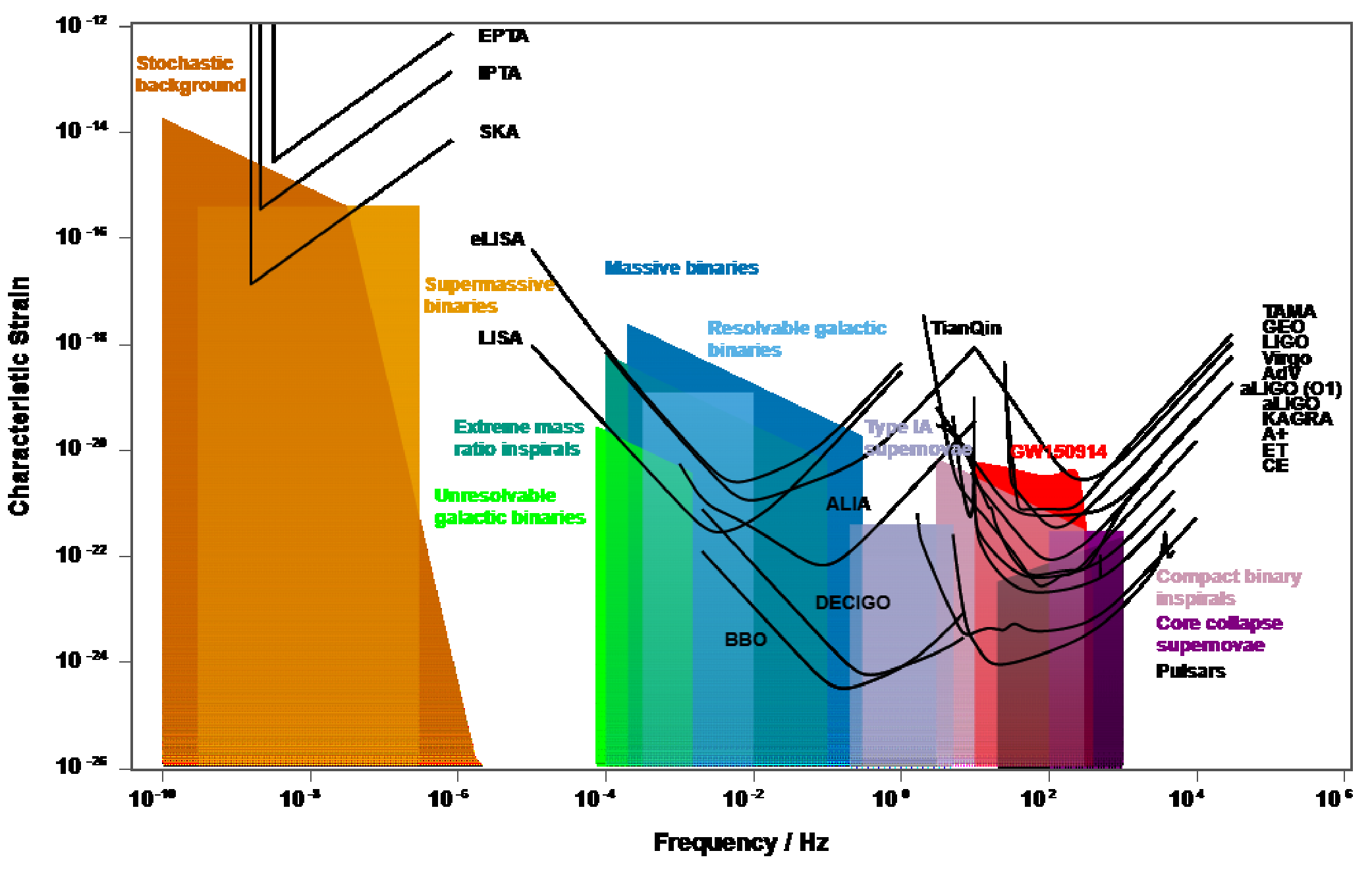}
	\caption{GW sensitivity curves of various detectors from several sources. The plot has been generated on-line using \url{http://gwplotter.com/} based on the Ref. \cite{Moore:2014lga}.}
	\label{fig:Fig11}
 \end{figure}
 The stochastic signatures of blue-tilted GW frequencies are important in probing deeply into the `pre-cosmic microwave background scenario' and therefore validating the paradigm of quintessential inflation, specifically the kination. Unfortunately, at present, no GW detectors are able to detect $\sim 10^{10}$ Hz ($f_{\mathrm{end}}$) frequency. It needs more sensitivity and sophistication. At least it is satisfactory that the frequencies $\sim 10$ Hz ($f_\mathrm{rad}$) are being detected in the ongoing experiments.\par There is an interesting connection between the GWs and the theory of inflation. Figure \ref{fig:Fig11} shows that every experimental group has observed GWs of strain $\sim 10^{-14}-10^{-24}$, which is undoubtedly too small. It is because, according to Eqs. (\ref{eq:H_end}, (\ref{eq:inf_scale}) and (\ref{eq:RD_strain}) the strain is proportional to tensor-to-scalar ratio $r$, which is very small ($<0.064$ \cite{Planck:2018jri}) for plateau-type model of inflation. In a previous work \cite{Sarkar:2023cpd} for the same potential considered here, the $r$ is found to be $\sim 10^{-4}-10^{-2}$ for $0.1\leq\alpha\leq 4.3$ at the pivot scale $k=0.002$ Mpc$^{-1}$. Therefore the observed tiny values of GW-strain indirectly support the framework of slow-roll inflation, just like Planck observation \cite{Planck:2018jri}. However, this aspect will be more prominent after the detection of $B$ modes with desired accuracy (see \cite{Sarkar:2021ird,Sarkar:2023cpd} for further discussions).
 \section{Summary and conclusions}
 \label{sec:conclusion}
In summary, we have
\begin{enumerate}
    \item constrained the values of $\alpha$ in the range $0.28\leq\alpha\leq 0.30$ using the expression of the scale of a new model of quintessential $\alpha$-attractor inflaton potential for two mass scales, \textit{viz.,} $M=4.05\times 10^{17}$ GeV and $\Tilde{M}=2.2\times 10^{10}$ GeV. The values of $\alpha$ are suitable for obtaining the required value of baryon-to-entropy ratio in the scenario of spontaneous baryogenesis (SB),
    \item studied SB by deriving the expressions of energy density and equation of motion of $\phi$ field, baryon excess and the freeze-out value of baryon-to-entropy ratio (BTER) at finite temperature by incorporating the derivative coupling of $\phi$ field with the non-conserving baryon current with a cut-off scale $M$. This coupling, describing a CP-odd effective interaction, brings extra terms in total field-energy density corresponding to the number densities of particles and anti-particles. By a calculation of statistical phase space distribution, the number density of particles is found to be greater than that of anti-particles for a non-zero value of field velocity, leading to the `baryon-excess' of the universe. This baryon-excess is translated into the computation of BTER. In this course, the effect of back-reaction is shown to be negligible in the dynamical equation of motion of the quintessential inflaton field,
    \item examined the evolution during kination of total field density $\rho_{\mathrm{kin}}$ and radiation density $\rho_{\mathrm{rad}}$ of the particles created through instant preheating (IP), with the expansion of the universe, which is measured by the number of e-folds $N$, elapsed after the end of inflation. The two energy densities are found to have same order of magnitude, which is very necessary for a successful BBN process during matter domination (which actually makes the IP a better reheating process rather than GPP). At $N = 12.1698$ there exists a cross-over between the above mentioned energy densities, which acts as a transition point between starting and ending of the SB process as well as between the domination of $\phi$-field and radiation,
     \item derived the expressions of the temperature, $T_{\mathrm{end}}$ at the end of inflation (or start of kination), the maximum thermalization temperature $T_{\mathrm{th}}$, and the temperature of radiation domination $T_{\mathrm{rad}}$, which have been used to examine the post-inflationary kination period related to SB,
    \item analyzed the evolution of these three temperatures and their respective number of e-folds with $\alpha$ within the range specified earlier. The order of magnitude of the values of $T_{\mathrm{th}}$ are close to that of $T_{\mathrm{end}}$ and too far from that of $T_{\mathrm{rad}}$ signifying the fact that, thermal equilibrium attains just after the end of inflation and long before the beginning of radiation era. This is needed to set a perfect stage for the BBN in matter dominated era; because, even a slight deviation of the order of the values of the said temperatures could be threatening for obtaining the observed constraints of BBN,
    \item considered an effective non-renormalizable $4$-fermion point interaction with a cut-off scale $\Tilde{M}$ and the associated rate of $B$-violation $\Gamma_{B-L}$ to compute the freeze-out temperature $T_F$ and the baryon-to-entropy ratio $\eta_F$ with the variations of $\alpha$ for fixed values of $M$ and $\Tilde{M}$ using the `freeze-out condition'. The obtained values of $T_F$ lie in between $T_{\mathrm{th}}$ and $T_{\mathrm{rad}}$ and much below the cut-off mass-scale $\Tilde{M}$, required for implementing the effective $4$-fermi construct method. The ratio $T_{\mathrm{th}}/T_{F}$ becomes consistent with the value given in the current literature and
    \item studied the variations of the amplitudes and the frequencies of the spectra of relic gravitational waves with $\alpha$ for two types of transitions - one is from inflation to kination ($KD$) and another is from $\phi$ domination to radiation domination ($RD$). The calculated magnitudes of the present-day RMS values of $RD-GW$ amplitudes, $\Omega_{GW,0}^{(\mathrm{RD})}$, are consistent with the sensitivity profile of the ongoing $GW$-detectors. The corresponding peak values, $(\Omega_{GW,0}^{(\mathrm{RD})})_{\mathrm{peak}}$, satisfy the BBN constraint quite satisfactorily. A blue tilted $KD-GW$ spectrum is also found to exist, which is an essential feature of SB.
\end{enumerate}
\par We, therefore, conclude that the $\alpha$ parameter of the new quintessential model of $\alpha$-attractor, considered in this paper, is found to be restricted within the range $0.28\leq\alpha\leq 0.30$. This range of $\alpha$ is just appropriate for explaining the baryon asymmetry at the GeV-scale and generation of `blue-spectrum' of relic gravitational waves. According to Ref. \cite{Sarkar:2023cpd}, this model is equally efficient in explaining early and late-time expansions of the universe for $0.1\leq\alpha\leq 4.3$. Thus, the combined results of the present paper and Ref. \cite{Sarkar:2023cpd} show that, the concerned potential is capable of unifying inflation, baryogenesis, quintessence and gravitational waves in a single framework. Also, it is quite interesting to notice that, fractional values of $\alpha$ have been found to be effective in order to obtain a viable model, which is consistent with the experimental constraints of all these four aspects.\par 
 Generically, as discussed in Section \ref{sec:kination}, the dynamics of the scalar field during kination and the associated parameters discussed in that section are all model independent. In this respect, from broader perspective the phenomenology of spontaneous baryogenesis is free from any particular choice of the inflaton potential. The corresponding results are universal for all types of models of inflation. But here, in the case of the model considered \cite{Sarkar:2023cpd}, we get an extra freedom to control the magnitude of field velocity \textit{vis-\`{a}-vis} the baryon-to-entropy ratio and that of other parameters. It is because, almost all the parameters depend upon the scale of inflation $V_{\mathrm{end}}$ and in the present model $V_{\mathrm{end}}$ is dependent on the parameter $\alpha$, instead of being a pure number (which is actually happened in most of the models) (see Eq. (\ref{eq:Vend})). Thus varying $\alpha$ we can in fact tune the magnitudes of all parameters which have the $\alpha$-dependencies through $V_{\mathrm{end}}$. This extra facility is only available for quintessential $\alpha$-attractor models, which can provide both a region of kination, required to study SB and a free parameter $\alpha$ to control the constraints of SB. Therefore the formalism developed in the present paper is only useful for quintessential $\alpha$-attractor models. Again, the range of $\alpha$ found, is specific for the particular form of the model used. It can be different for any other mathematical form of the model. So, the $\alpha$-parameter is not generically constrained, but, the method discussed is equally valid for all quintessential $\alpha$-attractor model. Thus the potential considered has a role in fixing the \textit{initial condition} of kination \textit{vis-\`{a}-vis} SB without changing the dynamics, which is model independent. \par The special provision we get in the present model is however not generic and can be considered as a limitation too, but so far as our motivation is concerned, it is an advantage. Because, following Refs. \cite{Sarkar:2021ird} and \cite{Sarkar:2023cpd} we aim at constraining the parameter $\alpha$ incorporating various constraints of physical phenomena like inflation, dark energy and baryon asymmetry. In the previous work \cite{Sarkar:2023cpd} we have already probed the inflationary slow-roll plateau and quintessential tail and found that $\alpha$ lies in the range $0.1\leq\alpha\leq 4.3$. Now, to probe the remaining section, the kination, we need a tool by which we can sneak into that regime, which is intrinsically model independent. This job is done by the special $\alpha$-dependent structure of the end-value of the potential (Eq. (\ref{eq:Vend})). We utilize this expression as an instrument to study the parameters of SB during kination. This however does not change the dynamics of the scalar field \textit{i.e.} the $a^{-3}$ dependency, it only shifts the its initial value $\Dot{\phi}_{\mathrm{end}}\approx \sqrt{2V_{\mathrm{end}}}$, which is helpful in obtaining the required value of BTER (see Eqs. (\ref{eq:BTER_2}) and (\ref{eq:Mod_eta_F})). It may be also pointed out that the non-renormalizable four-fermion point interaction used to derive the freeze-out temperature of the baryogenesis process actually hides the detailed particle physics scenario of renormalizable exchange and loop effects. These effects are not incorporated in the present formalism. \par The present studies therefore envisage a complete framework, in which the model considered is capable of unifying three regions by giving necessary constraints for inflation, baryogenesis and dark energy for $n=122$ and $\alpha\in[0.28,0.30]$.\par The fact that the parameter $\alpha$ is prone to acquire small and fractional values, as we move towards obtaining a generic model, can have an elementary connection with the origin of $\alpha$-attractor phenomenon from fundamental physics, such as supergravity and string theory. In the original $\alpha$-attractor formulation \cite{Kallosh:2013yoa} in $\mathcal{N}=1$ minimal supergravity, $\alpha$ represents the inverse curvature of $SU(1,1)/U(1)$ K\"{a}hler manifold. The underlying attractor nature is found to originate in supergravity from the radius $\sqrt{3\alpha}$ of the hyperbolic half-plane disk (called Poincar\'{e} disk) geometry \cite{Kallosh:2015zsa}, which has an interesting connection with the geometry of corresponding moduli space \cite{Carrasco:2015uma}. The hyperbolic geometry is further explored in details in maximal $\mathcal{N}=8$ supergravity \cite{Ferrara:2016fwe,Kallosh:2017ced} and found that, it is linked with the $7$-manifold with $G_2$ holonomy, where $\alpha$ can pick up discrete values, restricted within $3\alpha=1,~2,~3,\cdot\cdot\cdot,~7$. Interestingly, the values of $\alpha$, we found, are close to $3\alpha=1$. Some typical string theoretic models, like fibre inflation \cite{Cicoli:2008gp,Kallosh:2017wku} show that $\alpha =\frac{1}{2},~2$ can be obtained from type IIB flux compactification over elliptic $K3$ or $T^4$ fibration.\par In two recent works \cite{Let:2022fmu,Let:2023dtb} we developed a new way of deriving slow-roll inflaton potential in intersecting $D7$ branes configuration, from the stabilization of type IIB closed-string moduli of $T^6/Z_N$ type of Calabi-Yau threefold. Now, as a future work we wish to extrapolate the formulations developed in Ref. \cite{Sarkar:2023cpd} and the present paper in the paradigm of moduli stabilization in type IIB/F-theory with the following plans $-$
\begin{enumerate}
    \item Recently, we have derived in Ref. \cite{Sarkar:2024jxz} the actual (fractional) value of $\alpha$ \textit{vis-\`{a}-vis} the origin of $\alpha$-attractor behaviour within our framework \cite{Let:2022fmu,Let:2023dtb} by incorporating the effects of both open-string and closed-string moduli and compared it with other works, some of which are mentioned above. The inflaton field arises in the radial direction of the combination of open-string moduli. The resulting attractor potential manifests to have a generic slow-roll plateau and the universal predictions of cosmological parameters, which are two signature properties of $\alpha$-attractor potentials. The exact value of $\alpha$ is found to be $\alpha=\frac{1}{3}\approx 0.3$. Interestingly this obtained value of $\alpha$ lies within the range, which is derived in the present paper in the context of SB during kination. Again, as discussed above, this range of $\alpha$ is also compatible with the scenarios of inflation and DE.
    \item Therefore it is quite understandable that, there must be some connections among the string theoretic origin of $\alpha$ and the idea of quintessence and associated phenomena like SB and kination. In some contemporary literature \cite{Cicoli:2012tz,Cicoli:2018kdo,Brinkmann:2022oxy}, the fundamental source of quintessence/ dark energy from string theory and associated gravitational wave production \cite{ValeixoBento:2023qwt} have been studied. Now, it will be interesting to explore in our framework of moduli stabilization of type IIB string theory, a quintessential version of $\alpha$-attractor inflationary model (quite similar to as in Ref. \cite{Sarkar:2023cpd}), equipped with a generic slow-roll plateau as well as a generic quintessential tail and the associated fractional value of $\alpha$. This work will be valuable since, apart from having a provision to seek the route of the origin of dark energy or quintessence in $\alpha$-attractor phenomenon from string theory, we can also discover its effects on constraining $\alpha$ and generation of relic GWs. It will be challenging indeed so far as various string theoretic conjectures, types of compactification geometries and schemes of moduli stabilization are concerned.
    \item Attempts can also be made to connect the string motivated quintessential $\alpha$-attractor potential to the scenario of spontaneous baryogenesis (SB) by the same way as described in this paper and to another aspect \textit{viz}., the early dark energy (EDE) in the context of resolution of Hubble tension. Recently, in  Ref. \cite{Cicoli:2023qri} efforts have been made to realize the EDE in string theory. In Ref. \cite{Sarkar:2023vpn}, we have constructed a potential incorporating the EDE sector and found some interesting results. However, it is not clear, at present, whether the phenomena of SB and EDE are interconnected. Further studies are needed to unravel any possible connection.  
\end{enumerate}

\section*{Acknowledgments}
The authors acknowledge the University Grants Commission, The Government of India for the CAS-II program in the Department of Physics, The University of Burdwan. AS acknowledges The Government of West Bengal for granting him the Swami Vivekananda fellowship.
\appendix
\section{Density of \texorpdfstring{$\phi$}{} field with the effective interaction}
\label{app:A}
The Lagrangian of the $\phi$-field is given by
\begin{equation}
    \mathcal{L}_{\phi}=-\frac{1}{2}g^{\beta\gamma}\partial_{\beta}\phi\partial_{\gamma}\phi-V(\phi)-\frac{\lambda'}{M}\partial_{\beta}\phi J^{\beta}
\end{equation}
and the corresponding energy-momentum tensor can be calculated as
\begin{equation}
    \begin{split}
        T_{\nu}^{\mu}&=\delta_{\nu}^{\mu}\mathcal{L}_{\phi}-2g^{\mu\alpha}\frac{\partial\mathcal{L}_{\phi}}{\partial g^{\alpha\nu}}\\
        &=\delta_{\nu}^{\mu}\left(-\frac{1}{2}g^{\beta\gamma}\partial_{\beta}\phi\partial_{\gamma}\phi-V(\phi)-\frac{\lambda'}{M}\partial_{\beta}\phi J^{\beta}\right)+\left(g^{\mu\alpha}\partial_{\alpha}\phi\partial_{\nu}\phi+\frac{2\lambda'}{M}g^{\mu\alpha}\partial_{\alpha}\phi J_{\nu}\right)\\
        &=\delta_{\nu}^{\mu}\left(\frac{\Dot{\phi}^2}{2}-V(\phi)-\frac{\lambda'}{M}\Dot{\phi}J^0\right)+\left(g^{\mu 0}\Dot{\phi}\partial_{\nu}\phi +\frac{2\lambda'}{M}g^{\mu 0}\Dot{\phi}J_{\nu}\right).
    \end{split}
\end{equation}
The time-time component of $T_{\nu}^{\mu}$ is,
\begin{equation}
    \begin{split}
        T_0^0&=\left(\frac{\Dot{\phi}^2}{2}-V(\phi)-\frac{\lambda'}{M}\Dot{\phi}J^0\right)+\left(-\Dot{\phi}^2+\frac{2\lambda'}{M}\Dot{\phi}J^0\right)\\&=-\frac{\Dot{\phi}^2}{2}-V(\phi)+\frac{\lambda'}{M}\Dot{\phi}J^0,
    \end{split}
\end{equation}
from which we obtain the density as
\begin{equation}
\begin{split}
     \rho&=-T_0^0=\frac{\Dot{\phi}^2}{2}+V(\phi)-\frac{\lambda'}{M}\Dot{\phi}J^0\\
     &=\frac{\Dot{\phi}^2}{2}+V(\phi)-\frac{\lambda'}{M}\Dot{\phi}\left(n-\Bar{n}\right).
\end{split}
\end{equation}
\section{Derivation of baryon excess \texorpdfstring{$\Delta n$}{}}
\label{app:B}
The Fermi-Dirac distribution function for massless relativistic baryons having momentum $|\Vec{p}|\equiv p$, energy $\epsilon=p$ and chemical potential $\mu$ is,
\begin{equation}
        f(p,\mu)=\frac{1}{e^{\left(\frac{p-\mu}{T}\right)}+1}=\frac{e^{-p/T}}{e^{-\mu/T}+e^{-p/T}}.
\end{equation}
Under high temperature approximation we have $\mu/T\ll 1$ and terms $\mathcal{O}(\mu^2/T^2)$ are neglected. Therefore,
\begin{equation}
    \begin{split}
        &f(p,\mu)\approx \frac{e^{-p/T}}{1-\left(\frac{\mu}{T}-e^{-p/T}\right)}\\&=e^{-p/T}\left[1-\left(\frac{\mu}{T}-e^{-p/T}\right)\right]^{-1}\\
        &=e^{-p/T}\left[1+\sum_{\mathbb{Z}\ni n=1}^{\infty}\left(\frac{\mu}{T}-e^{-p/T}\right)^n\right]\\
        &=e^{-p/T}\Bigl[1+\sum_{\mathbb{Z}\ni n=1}^{\infty}\sum_{m=0}^{n} {}^nC_{m}\left(\frac{\mu}{T}\right)^{n-m}\left(-e^{-p/T}\right)^m\Bigr]\\
        &\approx e^{-p/T}\left[1+\sum_{\mathbb{Z}\ni n=1}^{\infty}{}^nC_{n-1}\left(\frac{\mu}{T}\right)\left(-e^{-p/T}\right)^{n-1}\right]\\
        &=e^{-p/T}+\frac{\mu}{T}\sum_{\mathbb{Z}\ni n=1}^{\infty}n\left(-1\right)^{n-1}\left(e^{-np/T}\right).
    \end{split}
\end{equation}
Similarly, the same distribution function for anti-baryons is
\begin{equation}
    f(p,-\mu)=e^{-p/T}-\frac{\mu}{T}\sum_{\mathbb{Z}\ni n=1}^{\infty}n\left(-1\right)^{n-1}\left(e^{-np/T}\right).
\end{equation}
Therefore,
\begin{equation}
    f(p,\mu)-f(p,-\mu)=\frac{2\mu}{T}\sum_{\mathbb{Z}\ni n=1}^{\infty}n\left(-1\right)^{n-1}\left(e^{-np/T}\right).
\end{equation}
Thus, the baryon excess,
\begin{equation}
\begin{split}
     &\Delta n = n-\Bar{n}=\int\Tilde{g}\frac{d^3\Vec{p}}{(2\pi)^3}\left[f(p,\mu)-f(p,-\mu)\right]\\
     &=\int_0^{\infty}\Tilde{g}\frac{4\pi p^2 dp}{(2\pi)^3}\left[f(p,\mu)-f(p,-\mu)\right]\\
    &=\Tilde{g}\frac{\mu}{\pi^2 T}\sum_{\mathbb{Z}\ni n=1}^{\infty}n(-1)^{n-1}\int_0^{\infty}p^2 e^{-np/T}dp.
\end{split}
\end{equation}
Putting $x=\frac{np}{T}$,
\begin{equation}
    \begin{split}
        &\Delta n=\Tilde{g}\frac{\mu T^2}{\pi^2}\sum_{\mathbb{Z}\ni n=1}^{\infty}\frac{(-1)^{n-1}}{n^2}\int_{0}^{\infty}x^2 e^{-x} dx\\
        &=\Tilde{g}\frac{\mu T^2}{\pi^2}\eta(2)\Gamma (3),
    \end{split}
\end{equation}
where $\eta (2)$ is the \emph{Dirichlet eta function}, which is related to \emph{Riemann zeta function} as $\eta(2)=\left(1-2^{1-2}\right)\zeta (2)$. Putting $\zeta (2)=\frac{\pi^2}{6}$ and $\Gamma (3)=2!$ we finally obtain
\begin{equation}
    \Delta n = \frac{\Tilde{g}\mu T^2}{6}=\frac{\Tilde{g}\lambda'\Dot{\phi}T^2}{6M}.
\end{equation}
\section{Equation of motion of \texorpdfstring{$\phi$}{} with the effective interaction}
\label{app:C}
We have the Lagrangian for the $\phi$ field
\begin{equation}
    \mathcal{L}_{\phi}=-\frac{1}{2}g^{\beta\gamma}\partial_{\beta}\phi\partial_{\gamma}\phi-V(\phi)-\frac{\lambda'}{M}\partial_{\beta}\phi J^{\beta},
\end{equation}
where the metric is composed of usual Friedmann metric components $g_{00}=-1$ and $g_{ii}=a^2$ such that $\sqrt{-g}=a^3$. Now, 
\begin{equation}
\begin{split}
    &\frac{\partial \mathcal{L}_{\phi}}{\partial (\partial_\mu \phi)}=-\frac{1}{2}g^{\beta\gamma}\left(\delta_{\beta}^{\mu}\partial_{\gamma}\phi+\partial_{\beta}\phi\delta_{\gamma}^{\mu}\right)-\frac{\lambda'}{M}\delta_{\beta}^{\mu}J^{\beta}\\
    &=-g^{\mu\beta}\partial_{\beta}\phi-\frac{\lambda'}{M}\delta_{\beta}^{\mu}J^{\beta}=-g^{\mu 0}\Dot{\phi}-\frac{\lambda'}{M}\delta_{0}^{\mu}J^0,
\end{split}
\end{equation}
\begin{equation}
    \begin{split}
        &\partial_{\mu}\left(\sqrt{-g}\frac{\partial \mathcal{L}_{\phi}}{\partial (\partial_\mu \phi)}\right)\\
        &=-\frac{\partial_\mu g}{2\sqrt{-g}}\frac{\partial \mathcal{L}_{\phi}}{\partial (\partial_\mu \phi)}+\sqrt{-g}\partial_\mu\left(\frac{\partial \mathcal{L}_\phi}{\partial\left(\partial_\mu \phi\right)}\right)\\
        &=\frac{6a^5\Dot{a}}{2a^3}\left(\Dot{\phi}-\frac{\lambda'}{M}J^0\right)+a^3\left(\Ddot{\phi}-\frac{\lambda'}{M}\partial_{0}J^0\right)\\
        &=a^3\left[\left(\Ddot{\phi}-\frac{\lambda'}{M}\partial_0 J^0\right)+3H\left(\Dot{\phi}-\frac{\lambda'}{M}J^0\right)\right]\\
        &=a^3\left[\left(\Ddot{\phi}-\frac{\Tilde{g}\lambda'^2T^2}{6M^2}\Ddot{\phi}\right)+3H\left(\Dot{\phi}-\frac{\Tilde{g}\lambda'^2T^2}{6M^2}\Dot{\phi}\right)\right]\\
        &=a^3\left[1-\frac{\Tilde{g}}{6}\left(\frac{\lambda'M_p}{M}\right)^2\left(\frac{T}{M_p}\right)^2\right]\left(\Ddot{\phi}+3H\Dot{\phi}\right)
    \end{split}
\end{equation} and
\begin{equation}
    \frac{\partial\left(\sqrt{-g}\mathcal{L}_{\phi}\right)}{\partial\phi}=-a^3\frac{\partial V}{\partial\phi}.
\end{equation}
Therefore, the equation of motion becomes
\begin{equation}
\begin{split}
    &\left[\partial_{\mu}\left(\frac{\partial}{\partial(\partial_{\mu}\phi)}\right)-\frac{\partial}{\partial\phi}\right]\sqrt{-g}\mathcal{L}_{\phi}=0\\
    \mathrm{or,}&\left[1-\frac{\Tilde{g}}{6}\left(\frac{\lambda'M_p}{M}\right)^2\left(\frac{T}{M_p}\right)^2\right]\left(\Ddot{\phi}+3H\Dot{\phi}\right)+\frac{\partial V}{\partial\phi}=0.
\end{split}
\end{equation}
\bibliography{biblio}{}

\end{document}